\documentclass[11pt,a4paper]{article}

\pdfoutput=1

\usepackage{jheppub}
\usepackage{amsmath}
\usepackage{amsfonts}
\usepackage{graphicx}
\usepackage{amssymb}
\usepackage{amssymb}
\usepackage{latexsym}
\usepackage{array}
\usepackage{bbold}

\newcommand{\asinh}{{\mathop{\mathrm{asinh}}}}

\usepackage{xcolor}
\definecolor{mygreen}{HTML}{006E28}

\newcommand{\fract}[2]{{\textstyle\frac{#1}{#2}}}

\title{Collisions of weakly-bound kinks in the Christ-Lee model}

\author[a,b]{Patrick Dorey,}
\author[c]{Anastasia Gorina,}
\author[d]{Tomasz Roma\'nczukiewicz}
\author[e]{and Yakov Shnir}

\affiliation[a]{Department of Mathematical Sciences, Durham University, UK}
\affiliation[b]{African Institute for Mathematical Sciences, Muizenberg, South Africa}
\affiliation[c]{Osmana\u{g}a Mahallesi Vahapbey Sokak No:8
Kadik\"oy, Istanbul 34714, Turkey}
\affiliation[d]{Faculty of Physics, Astronomy and Applied Computer Science, Jagiellonian University, Krak\'ow, Poland}
\affiliation[e]{Institute of Physics, University of Oldenburg,
Oldenburg D-26111, Germany}

\emailAdd{p.e.dorey@durham.ac.uk}
\emailAdd{nastya.gorina.2931@gmail.com}
\emailAdd{tomasz.romanczukiewicz@uj.edu.pl}
\emailAdd{shnir@maths.tcd.ie}

\date{April 2022}
\abstract{
We investigate soliton collisions a  one-parameter family of scalar field theories in 1+1 dimensions
which was first discussed by Christ and Lee \cite{Christ-Lee}. 
The models have a sextic potential with three local minima, and for suitably
small values of the parameter its kinks have an internal structure in the form of two weakly-bound subkinks. We show that for these values of the parameter kink collisions are best understood as an independent sequence of collisions of these subkinks, and 
 that a static mode analysis is not enough to explain resonant structures emerging in this model. We also emphasise the role of radiation and oscillon formation in the collision process.}
\begin{document}

\maketitle

\section{Introduction}
One of the most fascinating features of  soliton solutions in  non-integrable classical field theories 
\cite{Manton:2004tk,Vachaspati,Shnir2018} is the remarkable interplay between perturbative and non-perturbative dynamics, see for example \cite{kevrekidis2019dynamical}. It gives rise to an interesting  give-and-take mechanism of reversible energy exchange, which in particular results in the emergence of intricate resonance structures in collisions of solitons, such as kinks and antikinks in the 1+1 dimensional $\phi^4$ model   \cite{Campbell:1983xu,Goodman:2005,Makhankov:1978rg,Moshir:1981ja,Anninos:1991un}, and in the $\phi^6$ and other polynomial models \cite{Dorey:2011yw,Weigel:2013kwa,Gani:2014gxa,Demirkaya:2017euk,Khare:2014kva,Romanczukiewicz:2018gxb,Bazeia:2023qpf}. 

Numerical modeling of inelastic kink-antikink 
collisions reveal various mechanisms of the reversible energy transfer between the translational mode of the kink and its internal mode \cite{Campbell:1983xu,Anninos:1991un,Goodman:2005,Moshir:1981ja,Sugiyama:1979mi}; the energy can be also stored in excitations of scalar modes trapped by the kink-antikink pair \cite{Dorey:2011yw,Weigel:2013kwa,Gani:2014gxa}, or in
narrow quasi-normal modes \cite{Dorey:2017dsn,Campos:2019vzf}. Similar patterns have also been observed in resonant kink-impurity interactions \cite{Kivshar:1991zz,malomed2} and in boundary scattering of kinks  \cite{Arthur:2015mva,Dorey:2015sha,Lima:2018mzy}.

The collisional dynamics of the kinks becomes more involved as the number of modes localized on the kink increases \cite{Demirkaya:2017euk,Dorey:2021mdh}, or in the scattering of  wobbling kinks \cite{AlonsoIzquierdo:2020hyp}. The situation can be even more complicated in deformed models which support false vacua   \cite{Dorey:2019uap,Simas:2016hoo,Ashcroft:2016tgj,Gomes:2018heu,Adam:2019uat,Adam:2019prh}. 
Deformations of the potential may result in appearance of subkinks, which can be considered as weakly bound kinks interpolating between the true and false vacua \cite{sanati1999half,Mendonca:2015nka,Demirkaya:2017euk,Zhong:2019fub,Dorey:2021mdh}. 
In such cases, as we shall explore in the following, the collision of the solitons can be considered as a multi-step process, further complicated by the fact that
the initial impact of the forward pair of subkinks produces a burst of scalar radiation which deflects the trajectories of the second pair of subkinks and affects the outcome of the subsequent collisions\footnote{Discussions of how radiation pressure (positive and negative) influences the motion of kinks can be found in  \cite{Forgacs:2013oda,Forgacs:2008az,Romanczukiewicz:2003tn}. }.
Among other things this results in curious spine-like structures in plots of the final state of collisions as a function of the deformation parameter and initial velocity\,\cite{Dorey:2021mdh}; similar patterns can be seen in figure \ref{FullScan} below.

In our previous work \cite{Dorey:2021mdh} we briefly discussed this effect, alongside a plethora of other interesting phenomena, 
in the deformed sine-Gordon model of \cite{Dorey:2019uap} near the strong-deformation limit where it becomes
a $\phi^6$-like model with three degenerate vacua. In this paper we revisit this pattern in another model, first studied in a different context by Christ and Lee in 1975 \cite{Christ-Lee}. 

Our paper is structured as follows. In the next section we introduce the model and describe its static solutions and the fluctuations about these configurations. In section 3 we discuss various regimes of kink - antikink collisions. There special attention is paid to  collisions of weakly bound pairs of  subkinks and related radiation effects. Collisions between these subkinks and bions also play a role and these are given separate study in section 4.
We give our conclusions and further remarks in the final section.

\section{The model}
\begin{figure}
\centering
 \includegraphics[width=0.75\textwidth]{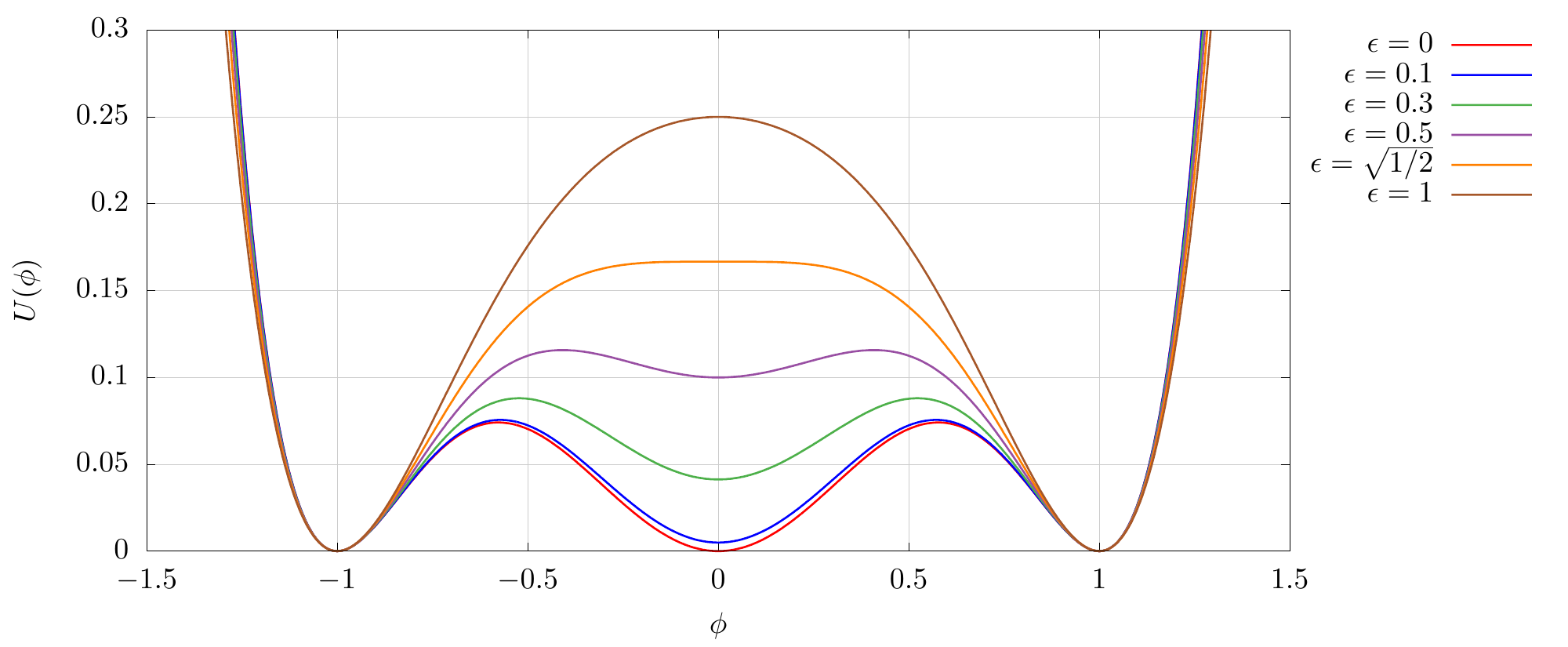}
\caption{The potential $U(\phi)$ of the model for different values of parameter $\epsilon$.}
\label{potentials}
\end{figure}

    We consider a model with the Lagrangian density that was discussed in \cite{Christ-Lee}
\begin{equation}
 \mathcal{L}=\frac{1}{2}\phi_t^2-\frac12\phi_x^2-U(\phi,\epsilon)\,,
\end{equation} 
where, setting the parameters $g$ and $\mu$ in \cite{Christ-Lee} to $1$ and $2$ respectively, the scalar potential is
\begin{equation}
 U(\phi,\epsilon) = \frac{1}{2(1+\epsilon^2)}(\epsilon^2+\phi^2)(1-\phi^2)^2\label{ft_potential}
\end{equation} 
Figure \ref{potentials} shows this potential for various values of $\epsilon$.

For $\epsilon=0$ the potential has three global minima at $\phi\in\{-1, 0, 1\}$
and
the model is the triply-degenerate $\phi^6$ model. Increasing $\epsilon$ breaks the degeneracy and
the vacuum at $\phi=0$ becomes false. 
For $\epsilon>1/\sqrt{2}$ this false vacuum is lost, and
the model has only two local minima; in the limit $\epsilon\to\infty$ it becomes the well-known $\phi^4$ model. 
The masses of small perturbations around the vacua are
\begin{equation}
    m_1=m_{-1}=2,\qquad m_0=\sqrt{\frac{1-2\epsilon^2}{1+\epsilon^2}}
\end{equation}

To discuss the limit $\epsilon\to 0$ it will sometimes be convenient to swap $\epsilon$ for an alternative parameter $\beta$, with
\begin{equation}
    \beta(\epsilon) = \log\left(1+\frac{1}{\epsilon}\right),\qquad \epsilon(\beta)=\frac{e^{-\beta}}{1-e^{-\beta}}\,.
\end{equation}
This maps $(0,\infty)\ni\epsilon\to\beta\in(\infty, 0)$.

\subsection{Static solutions}

\begin{figure}
\centering
\includegraphics[width=0.75\textwidth]{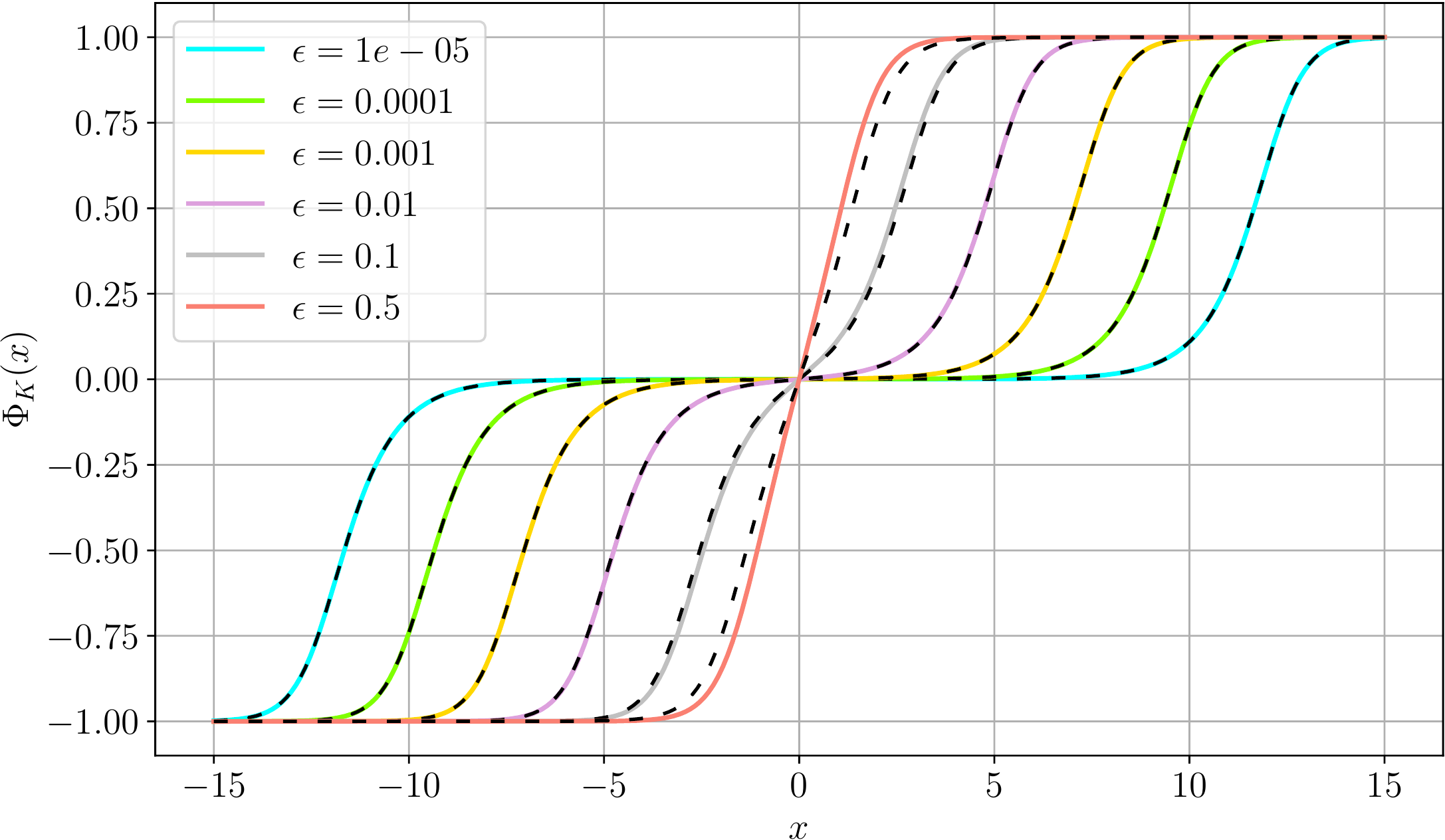}
\caption{Profiles of the stable static solutions. Dashed lines show
the approximation (\ref{apprx_profile}).}
\label{statics}
\end{figure}

For all $\epsilon>0$ the kink profile, interpolating between the vacua at $-1$ and $+1$,
is known analytically: 
\begin{equation}\label{large_kinks}
\Phi_K(x) = \frac{\epsilon \sinh(x)}{\sqrt{ 1 + \epsilon^2 \cosh^2(x)}}\,.
\end{equation}
Figure \ref{statics} plots some representative examples.
At $\epsilon=0$ the middle vacuum at $0$ becomes degenerate with the outer two and this solution is trivial.
Smaller $\phi^6$ kinks $\phi_{(-1,0)}$ and $\phi_{(0,1)}$ exist instead, interpolating between $-1$ and $0$, and  $0$ and $1$ respectively, where
\begin{equation}
    \phi_{(0,1)}(x)=\sqrt{\frac{1 + \tanh x}{2}} =\frac{1}{\sqrt{1+\exp(-2x)}}
\end{equation}
and $\phi_{(-1,0)}(x)=-\phi_{(0,1)}(-x)$. 
These smaller kinks emerge as
$\epsilon\to 0$ through the splitting of the $\Phi_K$ kink into two increasingly-separated $\phi^6$-like
subkinks, as can be seen by rewriting $\Phi_K$ as
\begin{equation}
\begin{split}
\label{split_kink}
  \Phi_K(x) &= \frac{\epsilon \exp(x)}{2\sqrt{ 1 + \epsilon^2 \cosh^2(x)}} - \frac{\epsilon \exp(-x)}{2\sqrt{ 1 + \epsilon^2 \cosh^2(x)}}\\[3pt]
  &=
  \phi_{\epsilon}(x-\overline x_0(\epsilon))-\phi_{\epsilon}(-(x-\overline x_0(\epsilon)))
\end{split}
\end{equation}
where 
\begin{equation}
\phi_{\epsilon}(x) = \frac{1}{\sqrt{(1+\fract{1}{4}\epsilon^2\exp(-2x))^2 + \exp(-2x)}}
\end{equation}
and
\begin{equation}\label{pos}
    \overline x_0(\epsilon) = \log\left(\frac{2}{\epsilon}\right)\approx \beta+\log(2).
\end{equation}
As $\epsilon\to 0$, $x_0$ grows,
$\phi_{\epsilon}(x)\to\phi_{(0,1)}(x)$ and the 
two subkinks emerge, 
positioned at $x=\pm \overline x_0$:
\begin{equation}\label{apprx_profile}
  \Phi_K(x) \sim  \phi_{(0,1)}(x-\overline x_0(\epsilon))-\phi_{(0,1)}(-(x-\overline x_0(\epsilon))).
\end{equation}
Alternatively, one can find the same expression by balancing the forces between the subkinks and the false vacuum as discussed, for example, in \cite{Dorey:2021mdh}. 

The mass of the full ($\epsilon>0$) kink is equal to
\begin{equation}
 M(\epsilon)=\frac{1}{2}-\frac{\epsilon^2}{4}+\frac{\epsilon^2(4+\epsilon^2)}{\sqrt{1+\epsilon^2}}\asinh(1/\epsilon)= \frac{1}{2} + \epsilon^2\log(\epsilon)+\mathcal{O}(\epsilon^2)
\end{equation}
Given that the mass of a single $\phi^6$ kink is $1/4$, this shows that for small values of $\epsilon$ the mass of the full kink exceeds the mass of two $\phi^6$ kinks by an amount 
$\epsilon^2\log(\epsilon)$, which can be attributed to the energy of the stretch of false vacuum lying between its two subkinks.

The presence of the false vacuum also allows for unstable infinite-energy static lump-like solutions, or sphalerons\footnote{The name is borrowed from \cite{Manton:1983nd, Klinkhamer:1984di}, where
saddle-point solutions of this type were  discussed in the context of the Weinberg-Salam Theory.}:

\begin{equation}
 \Phi_S(x) = \pm ab\sqrt{\frac{1-t^2}{b^2-a^2t^2}}\qquad\text{  where  }\qquad t=\tanh\frac{abx}{\sqrt{1+\epsilon^2}}
\end{equation}
and
\begin{equation}
 a = \sqrt{\frac{1}{2}\left(2-\epsilon^2-\epsilon\sqrt{4+\epsilon^2}\right)}, \qquad  b = \sqrt{\frac{1}{2}\left(2-\epsilon^2+\epsilon\sqrt{4+\epsilon^2}\right)}\,.
\end{equation}
Requiring $a$ to be real restricts $0<\epsilon<1/\sqrt{2}$ or $\beta>\log(1+\sqrt{2})\approx0.881374$. At the critical value $\epsilon=1/\sqrt{2}$ the false vacuum ceases to exist and the unstable lumps disappear.

\begin{figure}
\centering
\includegraphics[width=0.75\textwidth]{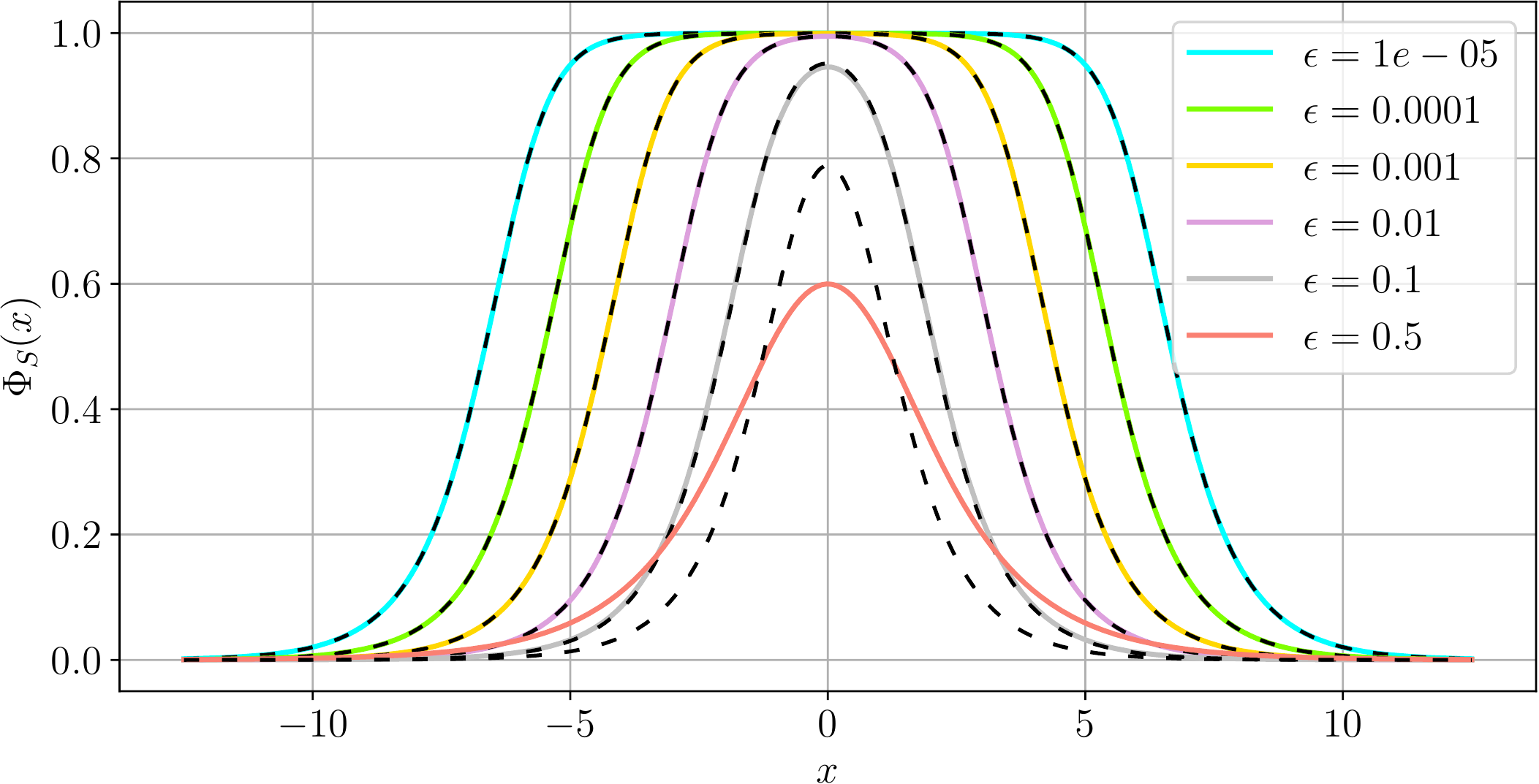}
\caption{Profiles of the unstable static solutions. Dashed lines correspond to the approximation (\ref{apprx_profile_s}) and (\ref{pos_s}).}
\label{statics_sph}
\end{figure}

For small values of $\epsilon$ the unstable lump  resembles a bound subkink -- antisubkink pair, submerged in the false vacuum which 
balances their mutual attraction:
\begin{equation}\label{apprx_profile_s}
    \Phi_S(x)\approx\sqrt{\frac{1+ \tanh (x+\overline x_1)}{2}}+\sqrt{\frac{1- \tanh (x-\overline x_1)}{2}}-1\,.
\end{equation}
The value of $\overline x_1$ can be found in a similar way as for the kink. The full formula is rather complicated, but the leading behaviour for small $\epsilon$ is
\begin{equation}\label{pos_s}
   \overline x_1(\epsilon)=\frac12\log\left(\frac{2}{\epsilon}\right)\approx\frac12\beta+\frac12\log(2).
\end{equation}
As illustrated in figure \ref{statics_sph}, this approximates the position very well for $\epsilon\lesssim 0.1$\,.

\subsection{Linearization}
\begin{figure}
\centering
 \includegraphics[width=0.75\textwidth]{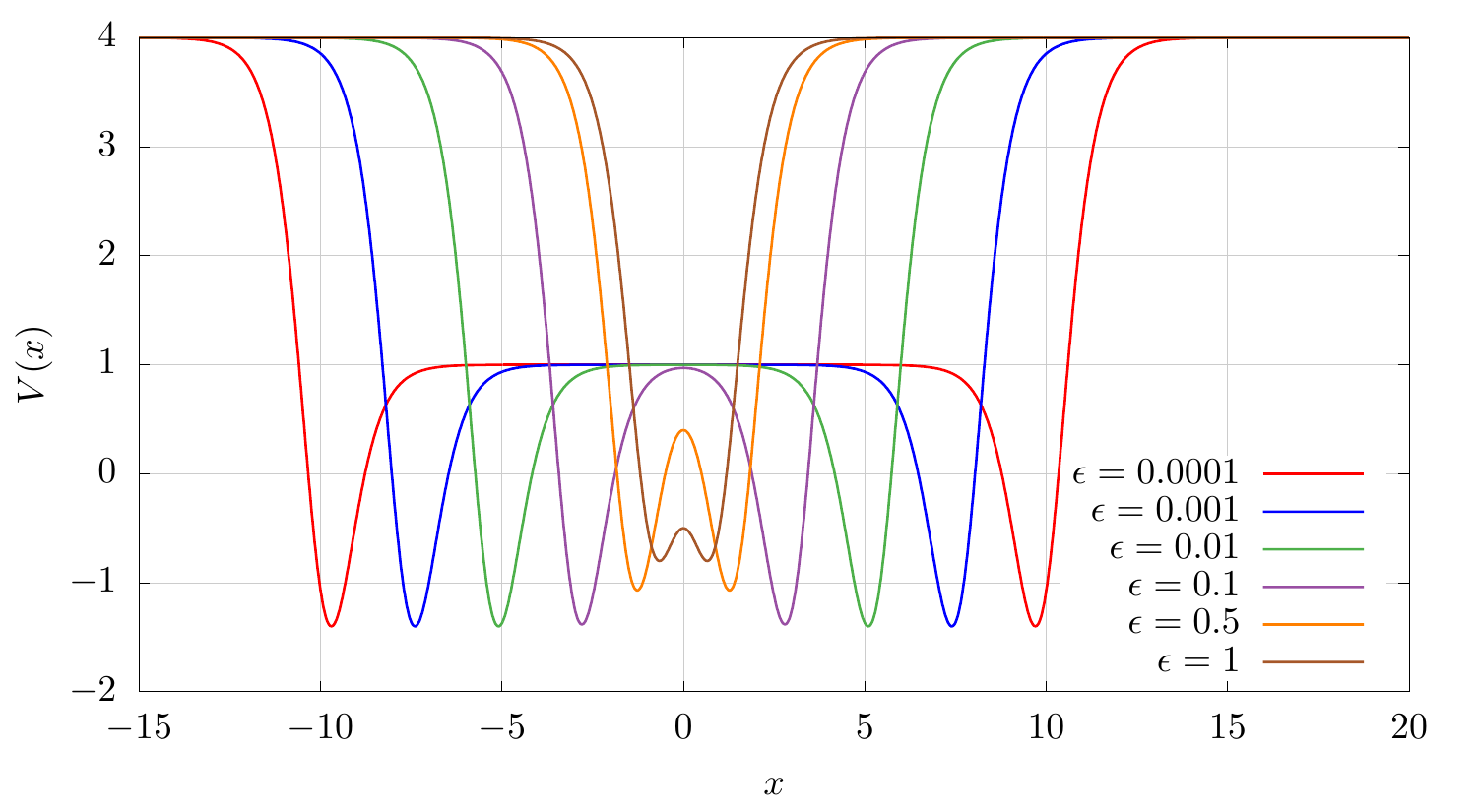}
\caption{Linearized fluctuation
potentials for the stable static solutions at different values of $\epsilon$.}
\label{linear_potentials}
\end{figure}

\begin{figure}
\centering
 \includegraphics[width=0.8\textwidth]{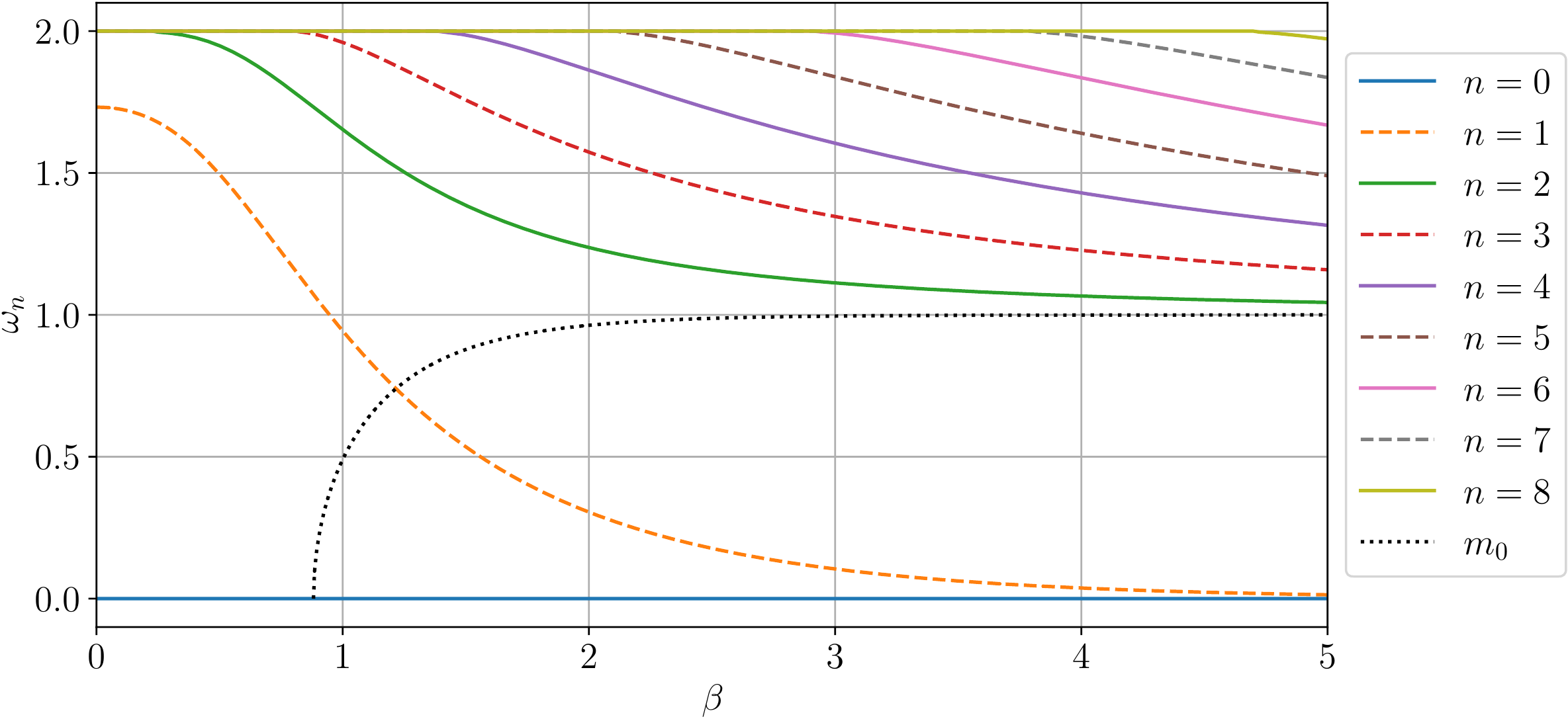}
\caption{Frequencies of the internal modes.}
\label{linear_frequencies}
\end{figure}

The linearized potential for fluctuations about a static solution $\phi$  is
\begin{equation}
 V(x) = \frac{1}{1+\epsilon^2}\left(15\phi^4+(6\epsilon^2-12)\phi^2-2\epsilon^2+1\right).
\end{equation} 
Figure \ref{linear_potentials} shows these potentials for
a single kink $\Phi_K(x)$ at various values of $\epsilon$, while figure~\ref{linear_frequencies} shows the frequencies 
of the corresponding bound modes
as a function of $\beta$. 
Due to the translational symmetry there is always a zero mode. 
For $\beta=0$ there is a single bound mode with frequency 
$\omega_1=\sqrt{3}$,
which reflects the fact that for $\beta=0$ 
the model is 
the $\phi^4$ model. For 
$\beta=0$ there is one more (nonintegrable) mode exactly at the threshold 
$\omega=2$ which becomes a proper bound mode for 
$\beta>0$. 
At values of $\beta$ equal to  $0.73289$, $0.98518$, $1.27641$, $1.61620$, 
$\ldots$ new modes appear below the threshold. 

As $\beta$ increases the frequencies of all modes decrease. The frequency of the first oscillating bound mode tends to $0$, while the frequencies of all other modes stay above the mass of the false vacuum $m_0(\epsilon)$. This is consistent with  the association of the lowest mode for large values of $\beta$ with translational oscillations of the weakly bounded subkinks.  The higher modes correspond to excitations trapped between the subkinks, with frequencies high enough to propagate in the false vacuum but still too low to propagate in the true vacuum.

\section{The small $\epsilon$ regime}

\begin{figure}
\centering
 \includegraphics[width=1\textwidth]{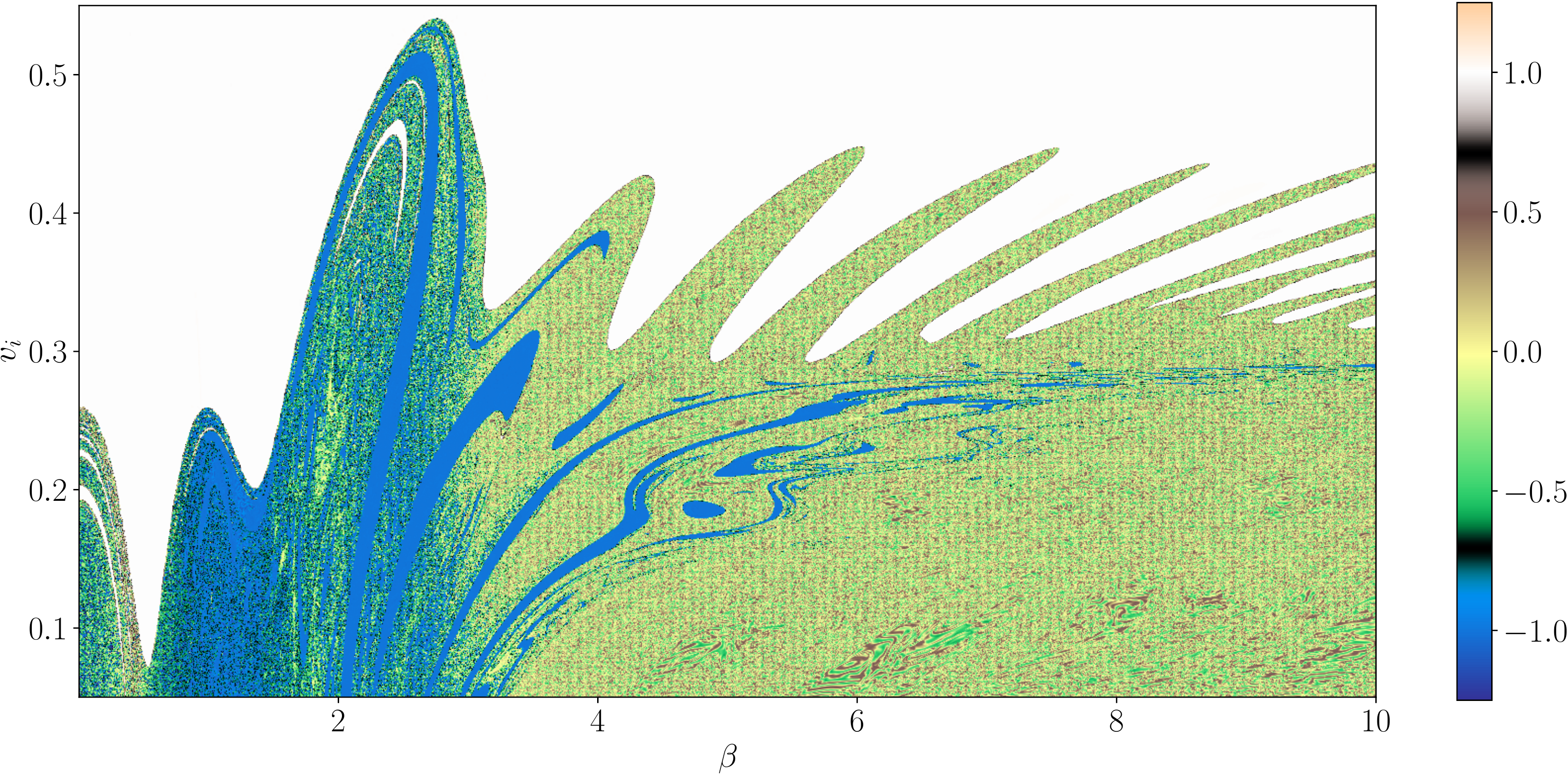}
\caption{Late-time field values $\phi(0,700)$ for $K \bar{K}$ collisions with initial velocities $\pm v_i$ in the Christ-Lee model. The pattern of spines for $\beta\gtrsim 4$ is similar to that found in a deformed sine-Gordon model in \cite{Dorey:2021mdh}.}
\label{FullScan}
\end{figure}

\begin{figure}
\centering
 \includegraphics[width=1\textwidth]{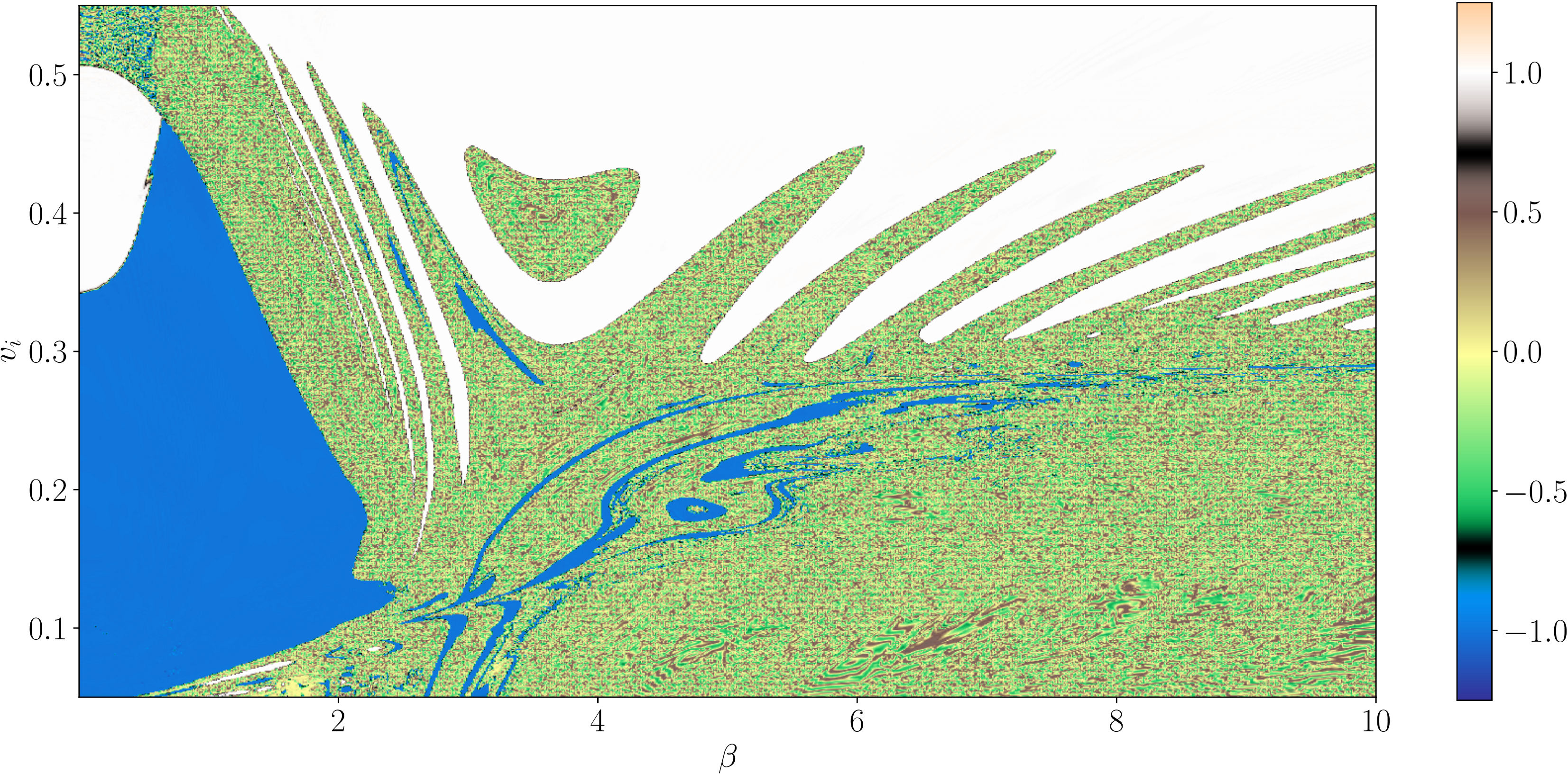}
\caption{Late-time field values $\phi(0,700)$ for the same Christ-Lee initial conditions as in figure \ref{FullScan}, but evolving the field using the $\phi^6$ ($\epsilon=0$) equation of motion.}
\label{FullScan_Phi6}
\end{figure}

For small values of $\epsilon$, or equivalently large values of $\beta$, it is useful to rewrite the
potential as
\begin{equation}
    U(\phi,\epsilon) = \frac{1}{2}\phi^2(\phi^2-1)^2-\frac{\epsilon^2}{2(1+\epsilon^2)}(\phi^2-1)^3.
\end{equation}
In this regime the field's evolution is almost entirely governed by the first term, which corresponds to the ordinary $\phi^6$ equation of motion. However, the $\phi^6$ model has very different static solutions than the $\epsilon>0$ Christ-Lee model. Thus we should expect that a collision between a kink and an antikink for $\epsilon>0$ would be very similar, not to a kink-antikink collision in the $\phi^6$ model, but rather to the collision of two $\phi^6$ kinks and two $\phi^6$ antikinks with suitably-tuned initial separations to match the particular value of $\epsilon$ under consideration. 
Figures \ref{FullScan} and \ref{FullScan_Phi6} test this idea.
Figure \ref{FullScan} 
shows the value of the field $\phi(0, T)$ after time $T=700$ for the Christ-Lee model with initial separation between the centres of the kink and antikink $2x_0=2\cdot 40$, and initial velocities $\pm v_i$.
Figure \ref{FullScan_Phi6} shows instead the value of $\phi(0, T)$ obtained using $\epsilon=0$ equation of motion, for the same initial conditions as figure \ref{FullScan}. Note that for $\epsilon=0$ the subkinks become independent  and the static kink solutions for any particular value of $\epsilon>0$ (comprising two weakly bound subkinks) are not static solutions for the $\epsilon=0$ equation of motion. But for sufficiently small values of $\epsilon$ the repulsion between subkinks can be neglected for the relatively short time scales over which the collisions take place. 

The match between the two scans for $\beta\gtrsim 5$
justifies the approximation, and indeed
beyond $\beta\approx10$ they become indistinguishable. Most notably the spine
structure is the same and the blue ``stream'' around $v_i\approx 0.289$ is clearly visible on both plots. We will argue that these features can be understood using the nontrivial internal structure of the kinks. 

\begin{figure}
\hspace*{0.02\textwidth}{\small a)}\hspace*{0.48\textwidth}{\small b)}\hspace*{0.41\textwidth}\\
\centering
 \includegraphics[width=0.49\textwidth]{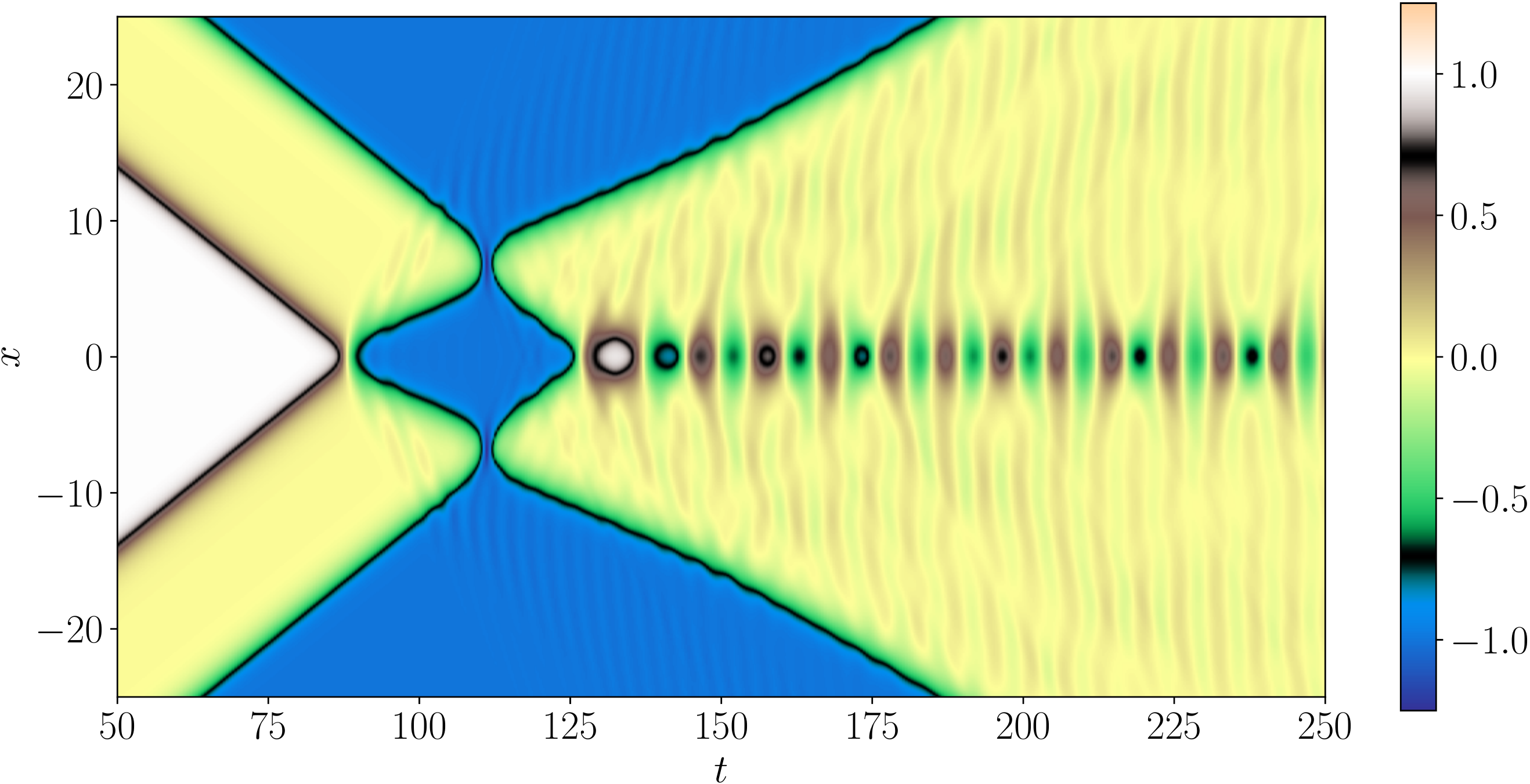}
 \includegraphics[width=0.49\textwidth]{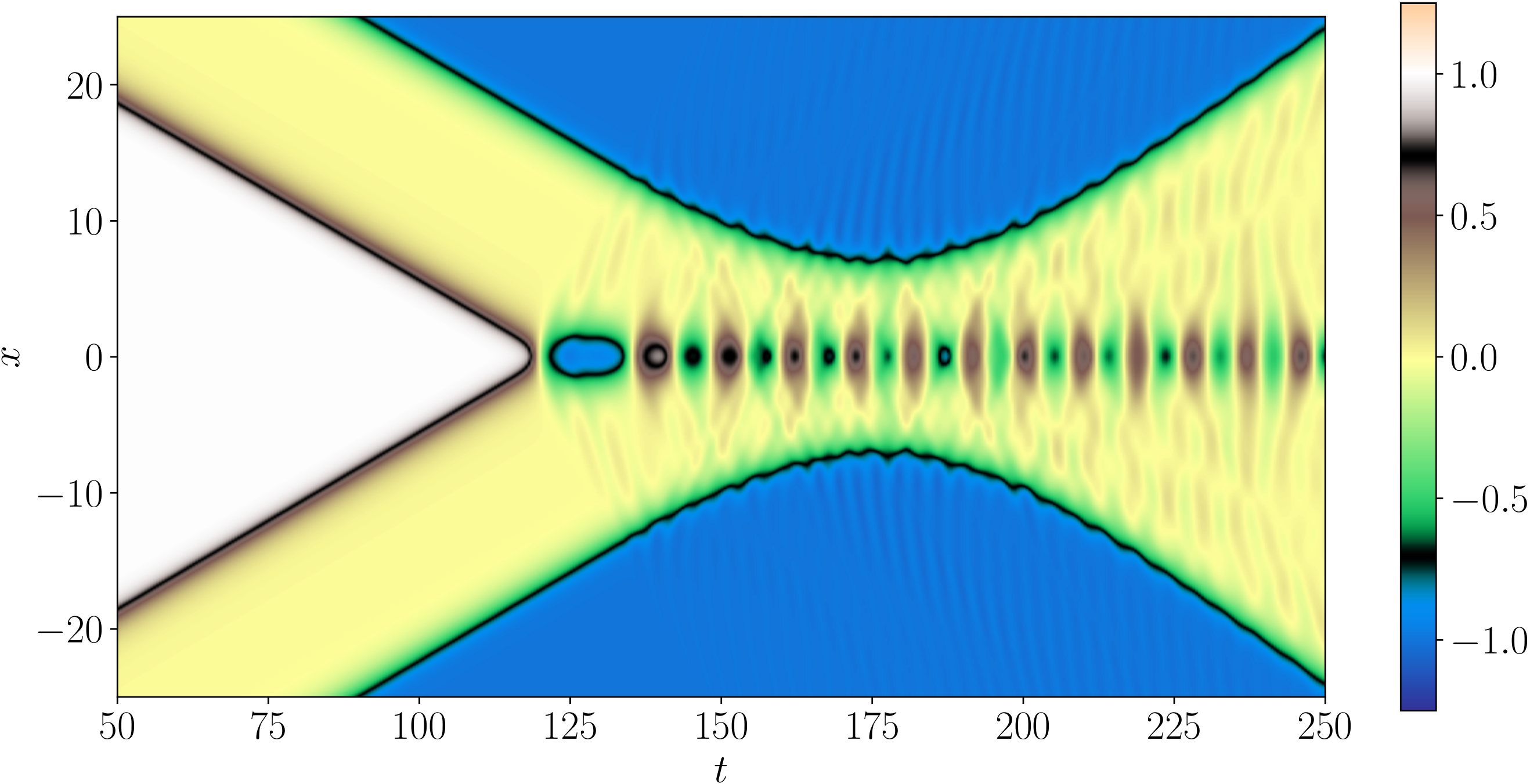}
\caption{Maps of two basic types of collision for large $\beta$. Examples for $\beta=8$ and $v_i=0.36$ (a) and $v_i=0.26$ (b).}
\label{CollisionTypes}
\end{figure}

To this end we first recall how the kinks and antikinks of the $\phi^6$ model scatter \cite{Dorey:2011yw}, bearing in mind that these correspond to subkinks of the Christ-Lee model. There are two distinct processes, depending on whether the vacuum at infinity is $0$ or $\pm 1$. The first, which we will call type I, corresponds to a
$(0,1)+(1,0)$ collision and results either in a single bounce for  $v_i>v_{cr}^I\approx 0.289$, or annihilation (with the formation of a centrally-located bion, seen also in figure \ref{CollisionTypes}b) for $v_i<v_{cr}^I$. By contrast, a $(1,0)+(0,1)$ collision, which we will call type II, can involve not only the direct formation of a bion but also a variety of resonant multi-bounce scenarios, leading to a  fractal-like structure for velocities between $0.0228<v_i<0.0457$. (For further work on the $\phi^6$ model see \cite{Weigel:2013kwa,Takyi:2016tnc,Gani:2014gxa} and also \cite{Adam:2022mmm}, where a detailed study of the collective coordinate description of the model was undertaken, confirming the role of the collective bound mode trapped between the $K\bar{K}$ pair in resonant type II scattering.)

Returning to the Christ-Lee model and looking at a $K \bar K$ collision for $\beta=8$ (figure \ref{CollisionTypes}a) we can indeed discern four subcollisions. The first, around $t=88$, is of type I. The critical value $v_i=v_{cr}^I=0.289$ for type I collisions
is visible on the scans shown in figures \ref{FullScan} and \ref{FullScan_Phi6} as the narrow (in $v_i$) blue ``streams' stretching over a wide range of $\beta$. 
In cases such as that shown in figure \ref{CollisionTypes}a  where the initial type I collision results in a bounce, two outgoing subkinks are produced which then collide with the two incoming outer subkinks (at  $t\approx 111$ in 
figure \ref{CollisionTypes}a). These collisions are topologically of type II, but since they take place in the presence of radiation escaping from the first collision we will denote them by II$^*$ with $^*$ indicating the additional radiation. Note that while, as mentioned above, pure type II collisions in the $\phi^6$ model have a fractal-like structure for velocities between $0.0228<v_i<0.0457$, in the current model the first collision results in escaping subkinks only for velocities larger than $0.289$, and the second collisions therefore take place with   velocities which are higher the the upper critical velocity for type II collisions.  In the absence of radiation, we would expect that the second collisions would always result in bounces. 
The fourth collision  ($t=127$) is of type I$^*$ because the subkinks after a previous bounce collide again. But this time the collision takes place in the presence of radiation from the earlier collisions, which can affect the outcome. As we will see shortly, the radiation can also affect the trajectories of the subkinks in between these collisions.

For velocities smaller than $v_{cr}^I$, a bion is formed after the first subcollision (figure \ref{CollisionTypes}b). Therefore instead of three subsequent subkink-subantikink subcollisions only a subcollision between a bion and two incoming subkinks takes place during the initial set of interactions.

In the following sections we will take a closer look at the new types of collisions and the role of background radiation.

\subsection{Spines as a result of radiation pressure}\label{section:spines}
In situations such as that shown in figure \ref{CollisionTypes}a, the first three collisions produce a large amount of radiation which affect the subsequent behaviour. This can be seen on all collision maps. In the $\phi^6$ model radiation exerts a pressure on solitons, pushing them in one direction only, no matter from which side of the soliton the radiation came \cite{Romanczukiewicz:2017hdu}. In fact, small fluctuations always act in such a way that the vacuum with smaller mass grows. In the current model this means that for large values of $\beta$ the radiation pressure  acts in the opposite direction to the vacuum pressure. Furthermore, at least for short time dynamics, the vacuum pressure can be neglected for large values of $\beta$. In contrast, we would expect that the last collision can be highly influenced by  radiation pressure.


\begin{figure}
\hspace*{0.02\textwidth}{\small a)}\hspace*{0.48\textwidth}{\small b)}\hspace*{0.41\textwidth}\\
\centering
\includegraphics[width=0.49\textwidth]{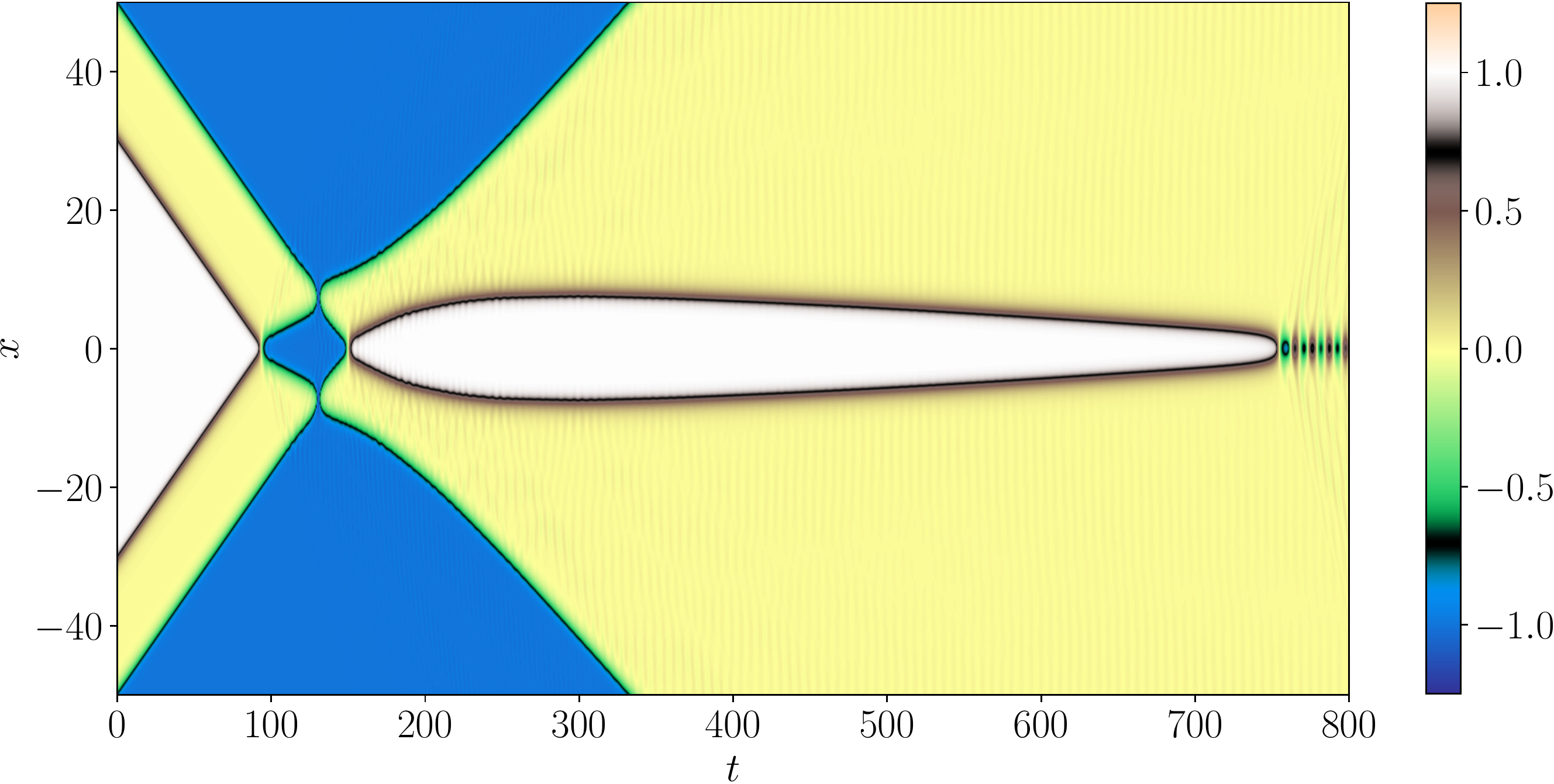}
 \includegraphics[width=0.49\textwidth]{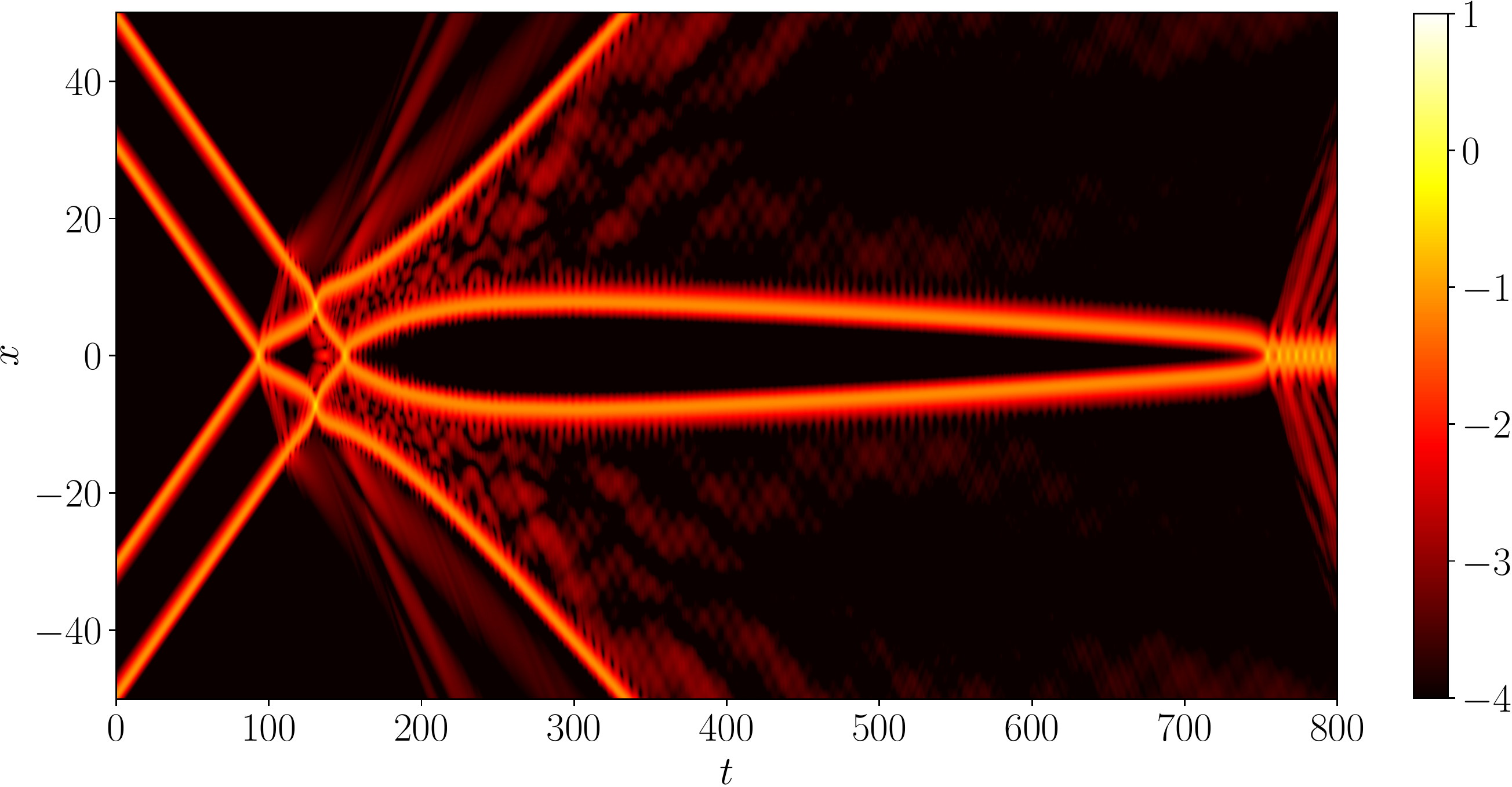}
\caption{A long false window (or false island) close to the base of a
spine, for $\beta=9.7600017$ and $v_i=0.31808$. In this and subsequent figures the left-hand plot shows the field values, while the right-hand plot shows the logarithm of the energy density.}
\label{LastCollision}
\end{figure}

Figure \ref{LastCollision} shows a collision for large $\beta=9.7600017$, very close to 
the edge of a spine.
After the second subkink collisions at $t\approx 130$
the outer subkinks bounce back, and then accelerate further after the collision. This is due to the background radiation which acts to expand $\phi=0$ vacuum. (Another possible cause of acceleration could be the presence of the inner kinks after the last collision, but they do not approach the outer kinks closely enough to interact efficiently.) 
After the fourth subkink collision at $t\approx 150$, the initially outgoing inner subkinks are deflected back towards the centre of collision, mostly by radiation.
This acceleration weakens with time and only when the subkinks approach each other to a separation of around $3$ or $4$ do they attract each other strongly and form a bion. This weakening acceleration can be explained by radiation escaping from the region so that only very weak fluctuations remain.

False vacuum pressure will ultimately cause the outer kinks to return to the central region, but over a very long time scale $T\sim e^{2\beta}$, well beyond the end of our simulation. This can be estimated as follows.
The false vacuum force is equal to the potential difference between the true and the false vacua, which is
\begin{equation}
    F=\Delta U=\frac{\epsilon^2}{2(1+\epsilon^2)}\leadsto\frac{1}{2}\epsilon^2\,.
\end{equation}
This approximately constant force turns back the subkink with initial velocity $v_0$ and forces it to recollide after time
\begin{equation}
    T=\frac{2v_0M}{F}\approx\frac{4v_0M}{\epsilon^2}=\frac{v_0}{\epsilon^2}\approx v_0e^{2\beta}.
    \label{recollide}
\end{equation}
This time can be even longer when radiation effects are taken into account.

Given the immense time scale needed for recollisions at these large values of $\beta$ and the even longer time bions need to decay, we stress that figures \ref{FullScan} and \ref{FullScan_Phi6} are only late-time approximations to the final state of the system. 
The soliton resolution conjecture \cite{tao2009solitons, Bizon:2021ose} implies that the final state consists only of solitons with a variety of velocities and dispersing radiation, and
the non-zero force between solitons in the model means that this state can contain at most one of these solitons can be static. A single static soliton would break the $x\to -x$ symmetry of our initial conditions, so all solitons in the final state must have non-zero velocity and ultimately leave the neighbourhood of the origin. Hence
the only possible limiting values of $\phi(0,t)$ as $t\to\infty$ are those of the true vacua, that is $\pm 1$, meaning that all points on our scans should be either white or blue if they were to show the limiting field values as $t\to\infty$.
As will be discussed in section 4 below, the bion-subkink re-collisions are chaotic and it can be hard to predict at which of these true vacua the field $\phi(0,t\to\infty)$ will end. 
However, from the energetic point of view full annihilation is more likely and therefore blue should dominate the lower parts of figures \ref{FullScan} and \ref{FullScan_Phi6}. 
Note that there is a clear distinction at the intermediate time scales shown in the figures between  
the yellow-green regions where a central and more slowly-decaying bion has been formed which is still in the process of decaying, and the regions
which are already blue, where energy has been exported from the central region by the emission of a pair of bions. These blue regions were called pseudowindows in our earlier paper \cite{Dorey:2021mdh}, and it would be interesting to give them a more mathematically precise definition.

\begin{figure}
\hspace*{0.02\textwidth}{\small a)}\hspace*{0.48\textwidth}{\small b)}\hspace*{0.41\textwidth}\\
\centering
 \includegraphics[width=0.49\textwidth]{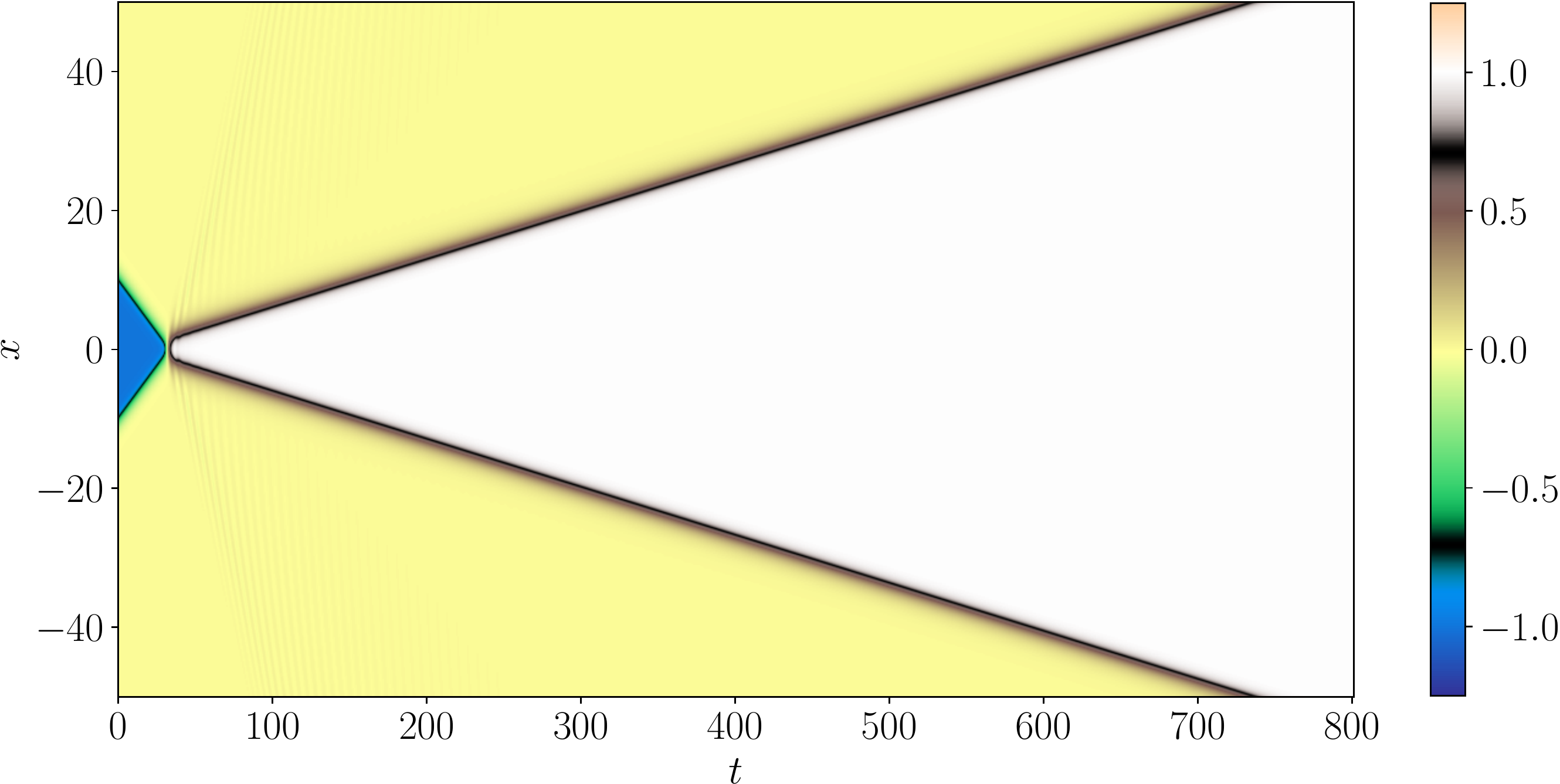}
 \includegraphics[width=0.49\textwidth]{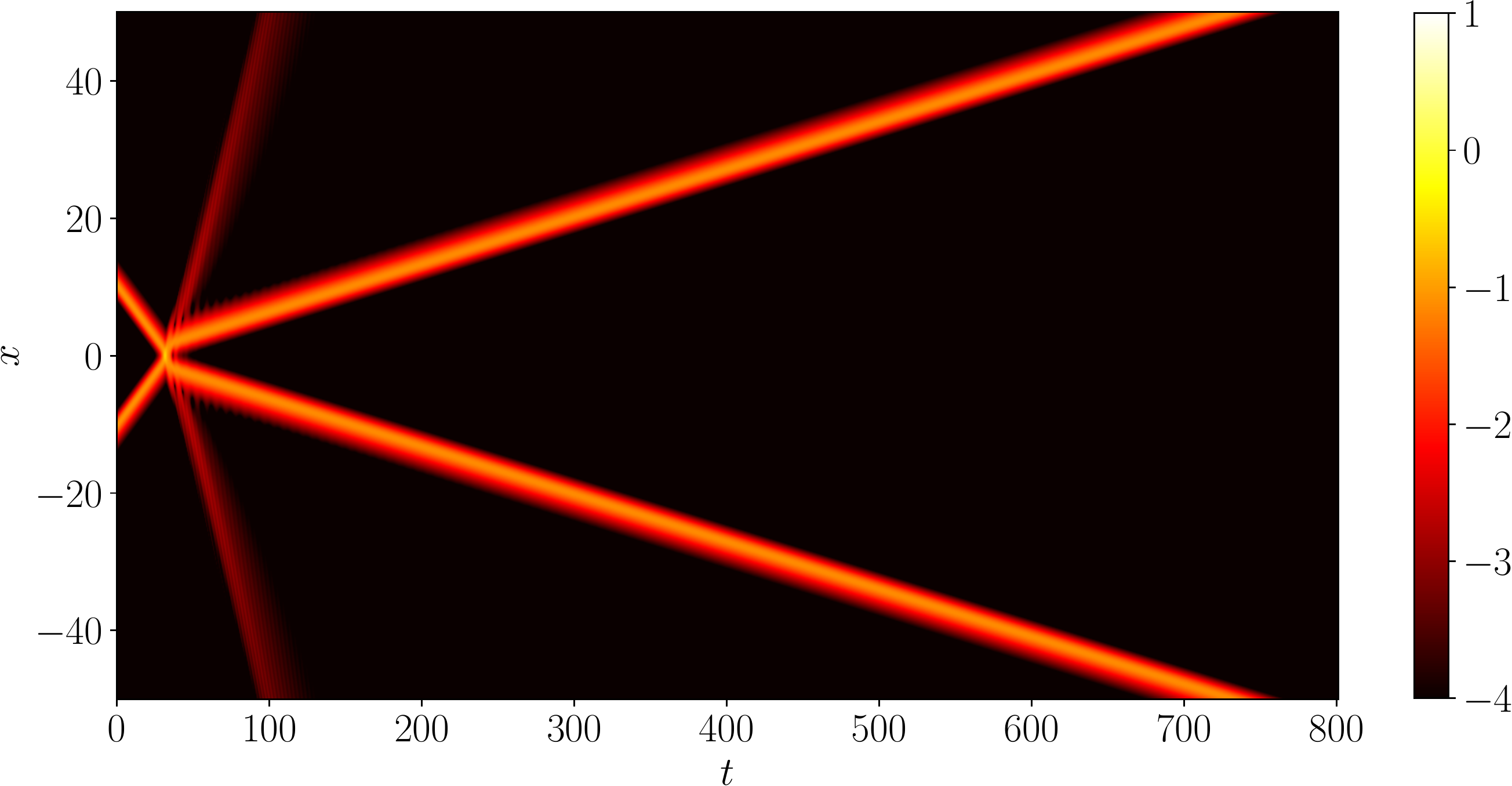}
\caption{A type I kink collision in the $\phi^6$ model with above-critical initial velocity $v_i=0.3$.}
\label{LastCollision_new2}
\end{figure}

\begin{figure}
\hspace*{0.02\textwidth}{\small a)}\hspace*{0.48\textwidth}{\small b)}\hspace*{0.41\textwidth}\\
\centering
 \includegraphics[width=0.49\textwidth]{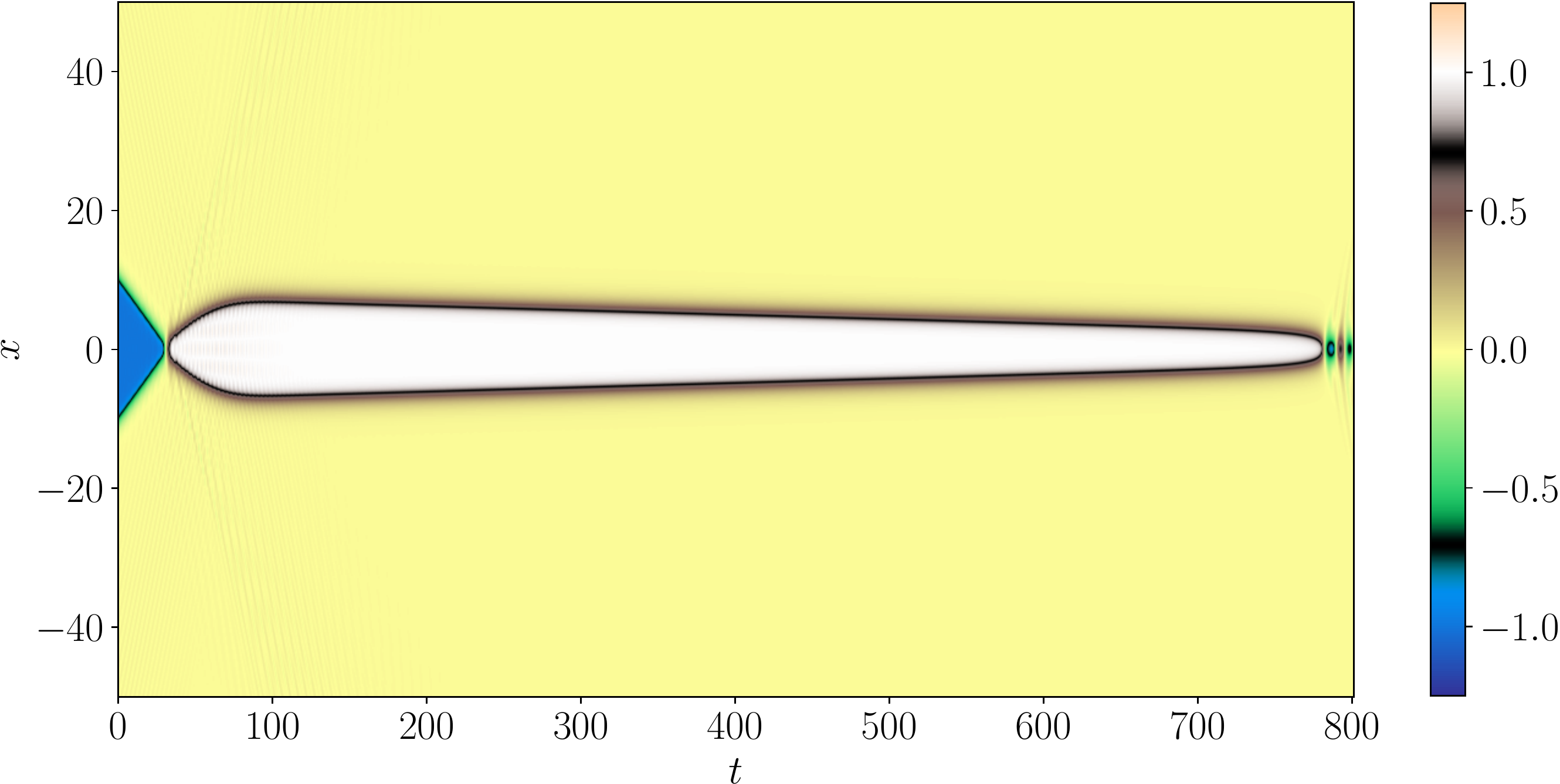}
 \includegraphics[width=0.49\textwidth]{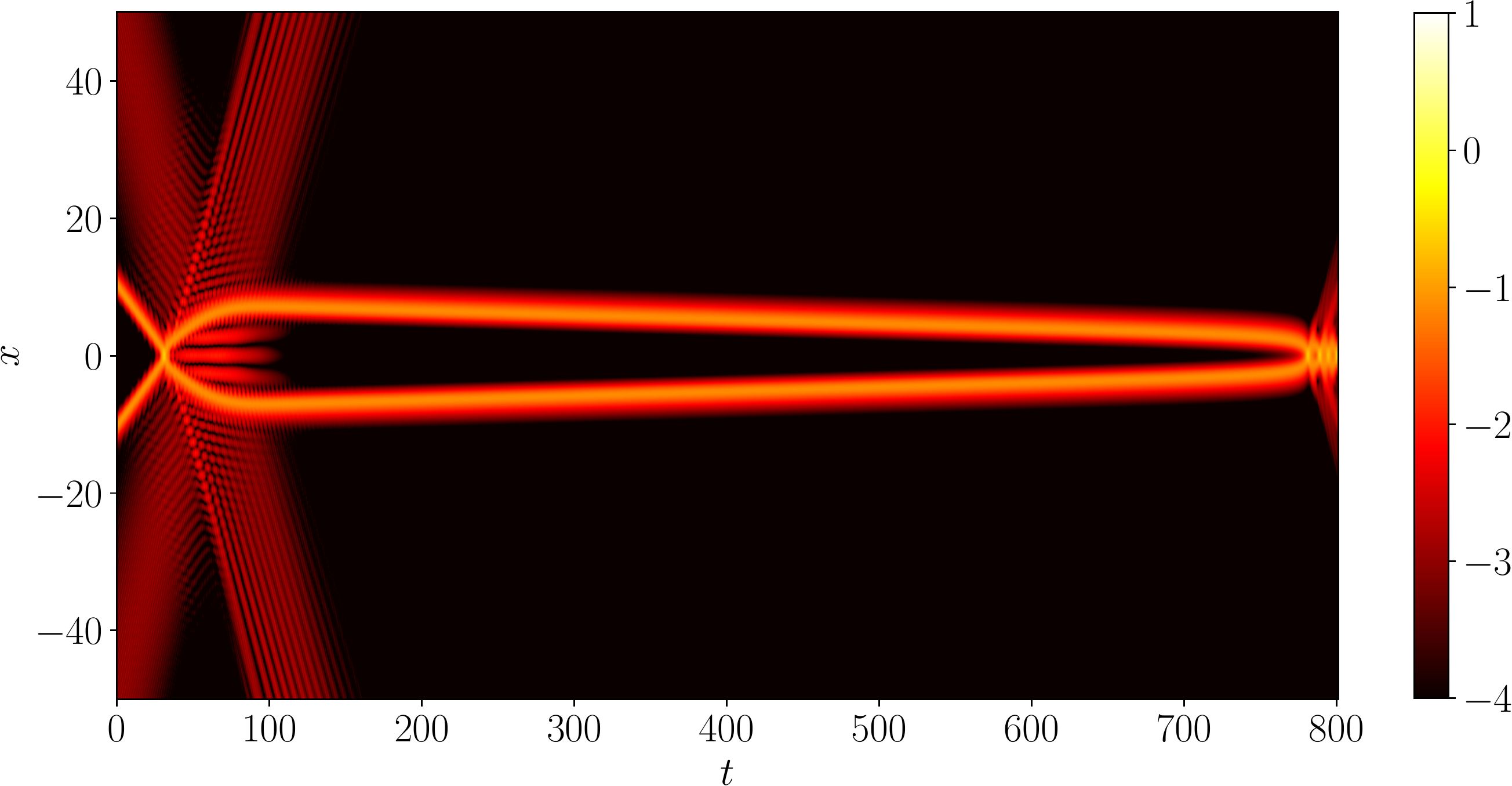}
\caption{The same collision as shown in figure \ref{LastCollision_new2}, with additional radiation in the form of two wavepackets with $\omega=2.2$, $s=50$, 
$\alpha=0$, $a=0.00125$, $A=0.028$ in the initial data.}
\label{LastCollision_new1}
\end{figure}

To assess the effect of perturbations such as radiation on subkink collisions,
we returned to the $\phi^6$ model and took initial data corresponding to a type I collision with and without an additional perturbing term, setting 
\begin{equation}
\phi(x,0)=\left[\phi_I(x,t)+\delta\phi(x,t)\right]_{t=0}\,,~~
\phi_t(x,0)=\left[\fract{\partial}{\partial t}(\phi_I(x,t)+\delta\phi(x,t))\right]_{t=0}
\end{equation}
with
\begin{equation}
\phi_I(x,t)=-\phi_{(0,1)}(\gamma(x-x_0-v_it)) - \phi_{(0,1)}(-\gamma(x+x_0+v_it))+1 
\end{equation}
being the initial data for a type I collision in the $\phi^6$ model, and $\delta\phi(x,t)$ being the possible perturbation.

Figure \ref{LastCollision_new2} shows the unperturbed ($\delta\phi=0$) collision for $v_i=0.3$, $x_0=10$, and figure
\ref{LastCollision_new1} shows the same collision with the addition of 
two travelling wave packets of amplitudes $A$ and widths $1/\sqrt{a}$ into the initial data:
\begin{equation}
    \delta \phi(x, t) = Ae^{-a(x+vt-s)^2}\sin(\omega t+k(x{-}s)+\alpha) + Ae^{-a(x-vt+s)^2}\sin(\omega t-k(x{+}s)+\alpha)
    \label{wavepackets}
\end{equation}
where $\omega=\sqrt{k^2+1}$,
$\pm s$ are the locations of the wavepackets at $t=0$,  
$\mp v=\mp d\omega/dk$ are their group velocities, 
and $\alpha$ is a phase.
The wave packets reach the central region just after the initial collision and push the outgoing subkinks back towards the centre, but after passing by the acceleration vanishes. This models quite well the trajectories of the inner subkinks after the last collision shown in figure \ref{LastCollision}. 

\begin{figure}
\hspace*{0.02\textwidth}{\small a)}\hspace*{0.48\textwidth}{\small b)}\hspace*{0.41\textwidth}\\
\centering
\includegraphics[width=0.49\textwidth]{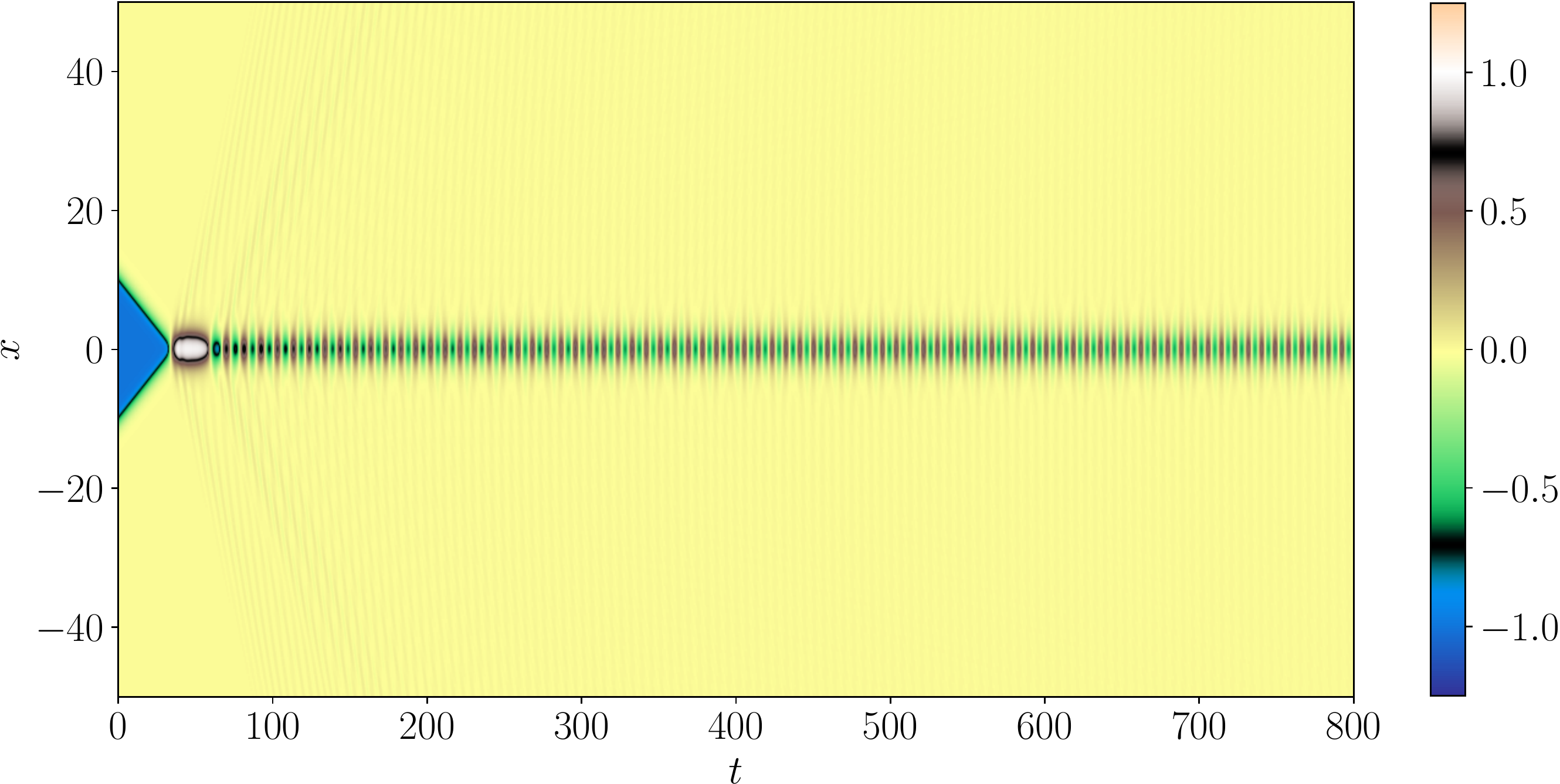}
\includegraphics[width=0.49\textwidth]{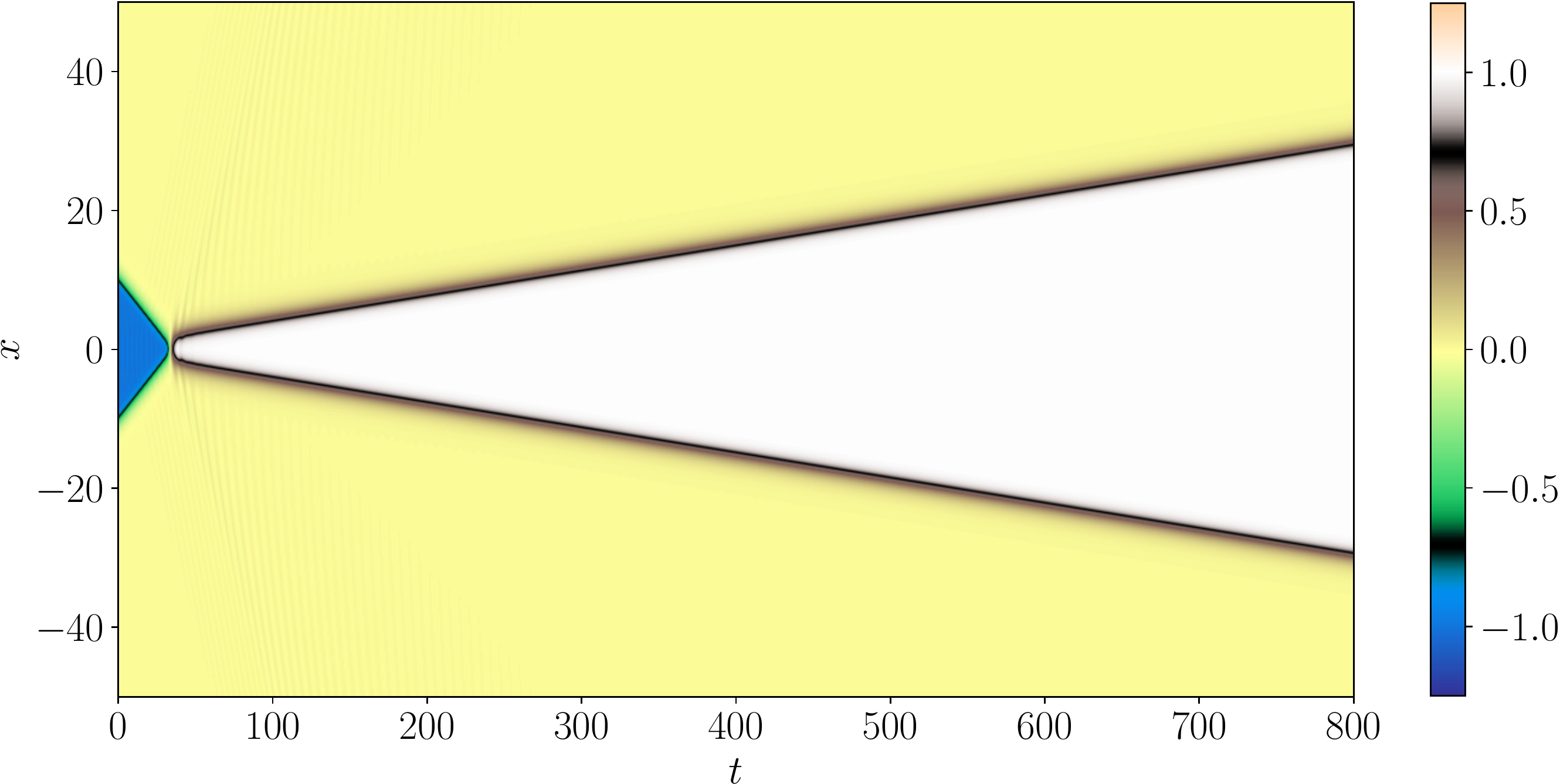}
\caption{(a) A $\phi^6$ type I collision just below the critical velocity for $v_i=0.28$ and $x_0=10$, (b) the same collision but with a gaussian perturbation at centre $\delta\phi=-0.028e^{-0.1x^2}$ for $v_i=0.28$.}
\label{LastCollision2}
\end{figure}

Similar results can be found when instead of moving wave packets the initial simulation starts with a gaussian perturbation at the centre (figure \ref{LastCollision2})
\begin{equation}
    \delta\phi = -Ae^{-bx^2}.
\end{equation}
For the example shown we again set $x_0=10$, but took $v_i= 0.28$, which is below the critical velocity,  and chose $A=0.028$ and $b=0.1$.   

 In the simulations shown in figures \ref{CollisionTypes} and \ref{LastCollision}\,a we see that the first collision generates outgoing  waves which then reflect from the outer subkinks before the second and the last collisions and turn back as wave packets. Some residual radiation also stays near the centre as low frequency slowly moving waves. Therefore the overall effect is rather a superposition of an almost stationary wave and wave packets coming from a distance.

\begin{figure}
\hspace*{0.02\textwidth}{\small a)}\hspace*{0.48\textwidth}{\small b)}\hspace*{0.41\textwidth}\\
\centering
\includegraphics[width=0.49\textwidth]{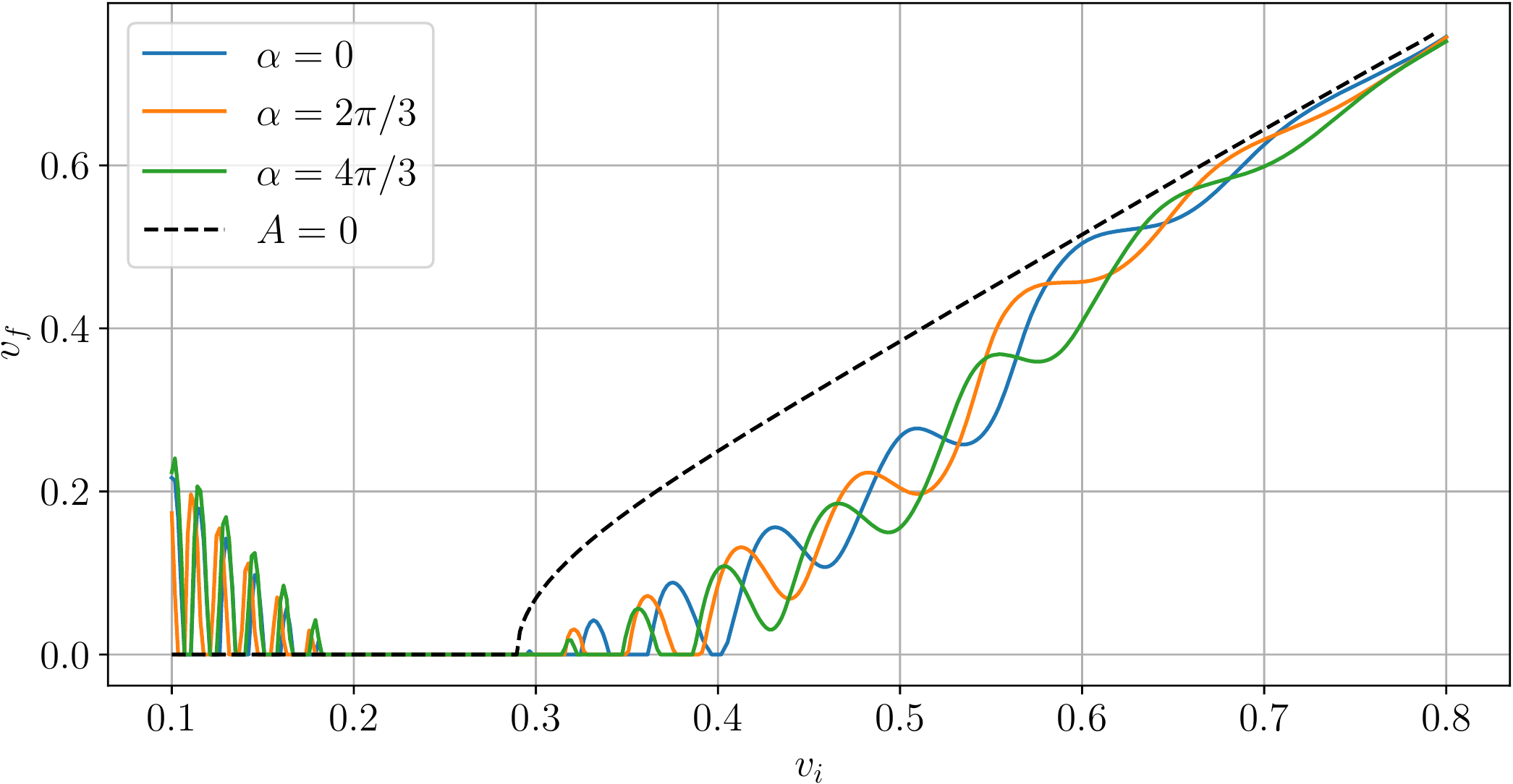}
\includegraphics[width=0.49\textwidth]{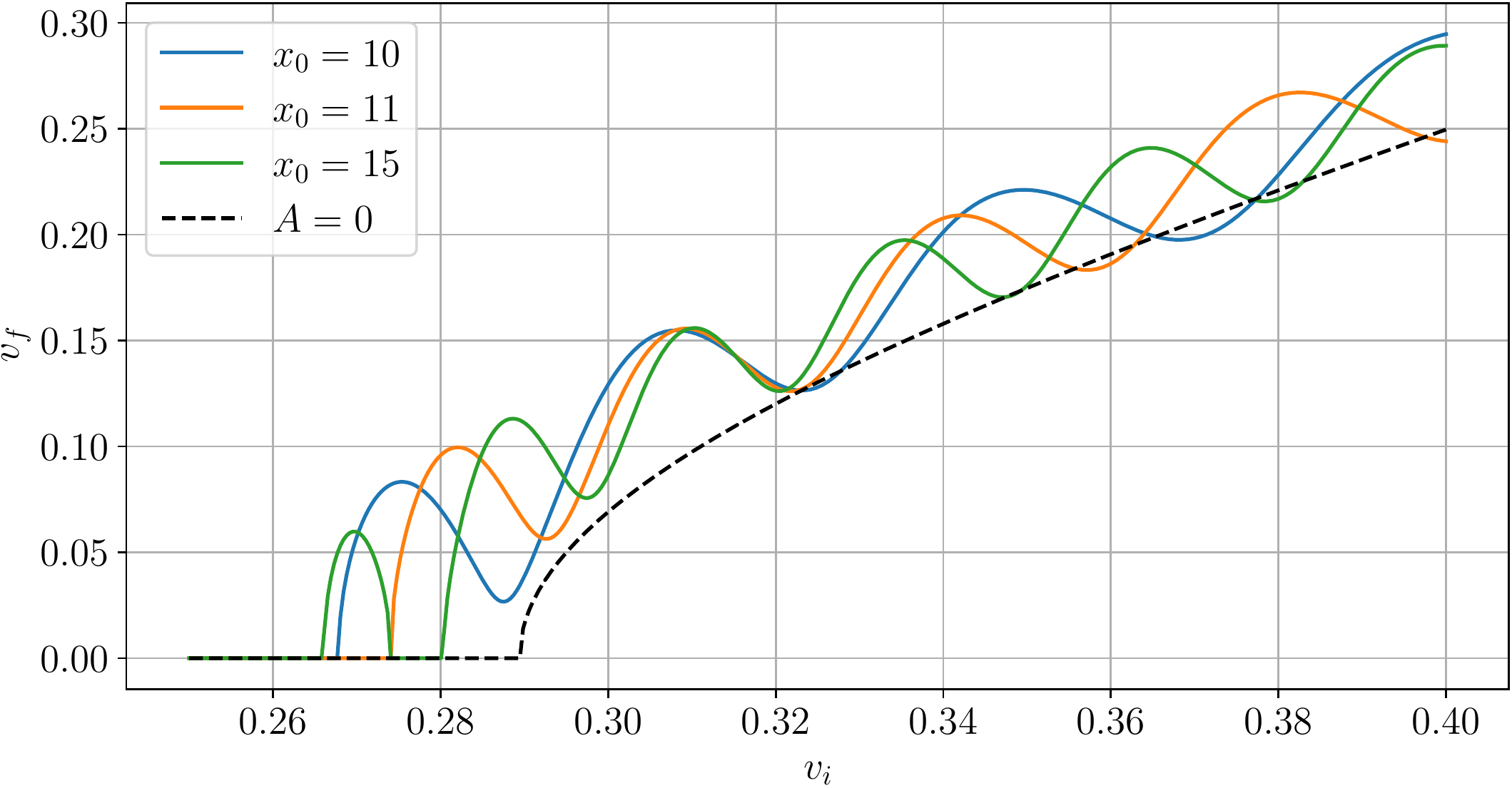}
\caption{Final velocities for type I collisions of two $\phi^6$ kinks with perturbed initial data. Plot~(a) shows the effect of incoming wave packets with $\omega=2.2$, $A=0.028$, $s=50$ for different phases $\alpha$, while plot
(b) shows the effect of a gaussian perturbation at the centre with $A=0.028$, $b=0.1$ for different initial locations $\pm x_0$ of the $\phi^6$ kinks. In both cases the $A=0$ curve shows the unperturbed situation.}
\label{FinalVelocity}
\end{figure}

\begin{figure}
\centering
\includegraphics[width=0.75\textwidth]{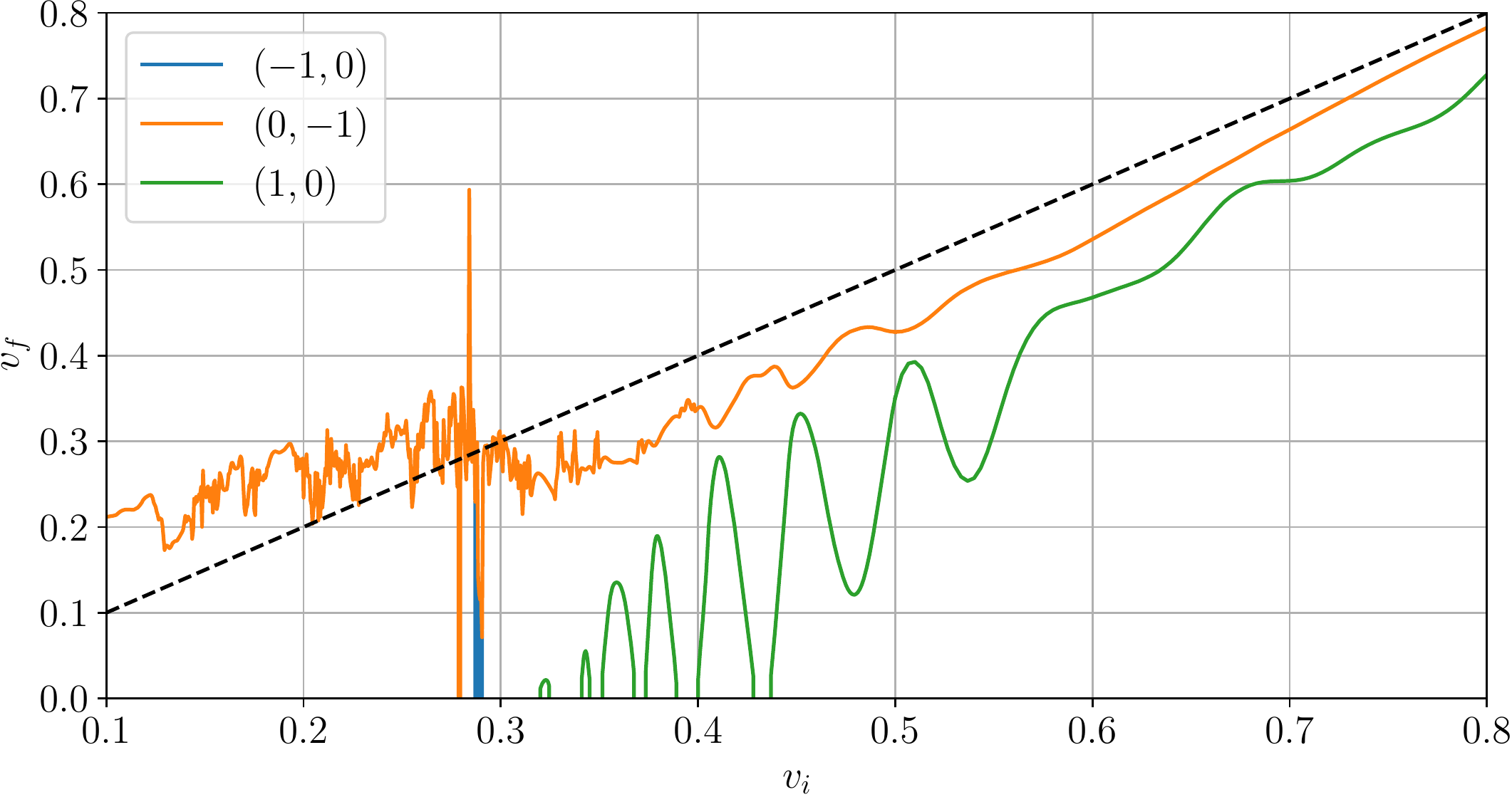}
\caption{Final velocity of subkinks as a result of the collision for Christ-Lee model for $\beta=10$. For $v_i<0.28$ the gain in velocity is due to the radiation. 
}
\label{FinalVelocities}
\end{figure}

Moreover, the force exerted by the waves is not constant on short time scales, and the phase of the wave packet can influence the outcome of the evolution (c.f.\ figure \ref{FinalVelocity}). Small differences in the phase can result in ultimate annihilation of the inner subkinks or their escape. 
We think that exactly this mechanism is responsible for the creation of spines. 
Velocities of individual subkinks measured after the collision plotted in figure \ref{FinalVelocities} show great similarity to figure \ref{FinalVelocity}.

At this point it is worth mentioning that double kink collisions were also extensively studied in the double sine-Gordon model \cite{CAMPBELL1986165, Campos:2021mkn, Gani:2017yla}. However, the double sine-Gordon model interpolates between two versions of the integrable sine-Gordon model, as a certain parameter $R$, corresponding to half the separation between the subkinks, varies from $0$ to $\infty$ \cite{Campos:2021mkn}. One of the consequences of integrability is the fact that solitons do not lose energy during  collisions, which translates into the vanishing of the critical velocity both at $R=0$ and also as $R\to\infty$.  This is one reason why spines are not observed in the double sine-Gordon model for collisions of unexcited kinks, even in the large $R$ limit.

\subsection{Further radiation effects}
At large values of $\beta$ the two subkinks making up a full kink are weakly bound, separated by a large region of false vacuum. The vacuum pressure trying to collapse this region is very small in comparison with other effects such as radiation pressure, and can be ignored over typical collision timescales. As evidenced by the match between figures \ref{FullScan} and \ref{FullScan_Phi6} for $\beta\gtrsim 5$, the motion of subkinks in this regime is very close to that of  full kinks in the $\phi^6$ model.
Radiation pressure in the $\phi^6$ model was studied in \cite{Romanczukiewicz:2017hdu}, where it was shown that both negative and positive radiation pressure can be exerted on a single $\phi^6$ kink. Since the vacuum at $\phi=0$ has the lower mass, radiation coming from the vacua at $\phi=\pm 1$ to that at $\phi=0$ causes negative radiation pressure, while radiation moving from the $\phi=0$ vacuum to those at $\phi=\pm1$ causes positive radiation pressure. In other words radiation in the $\phi^6$ model
always pushes subkinks so as to expand the region of $\phi=0$ vacuum at the expense of $\phi=\pm 1$ vacua. 

\begin{figure}
\centering
    \includegraphics[width=\textwidth]{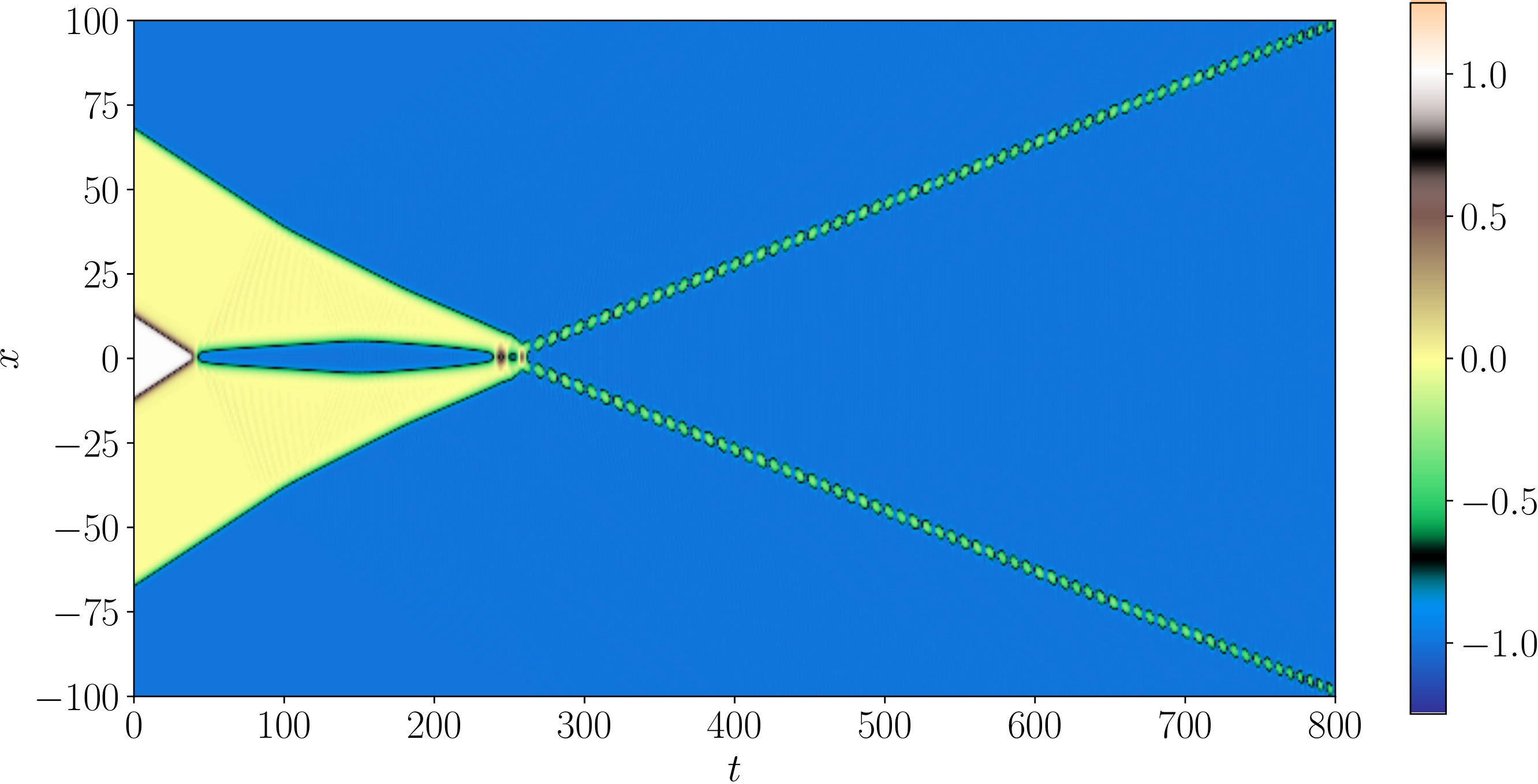}
\caption{Field values for a  kink collision at large $\beta$: \mbox{$\beta=28.199532$}, $v_i=0.291247$.}
\label{Radiation_effects1}
\end{figure}

\begin{figure}
\centering
\includegraphics[width=\textwidth]{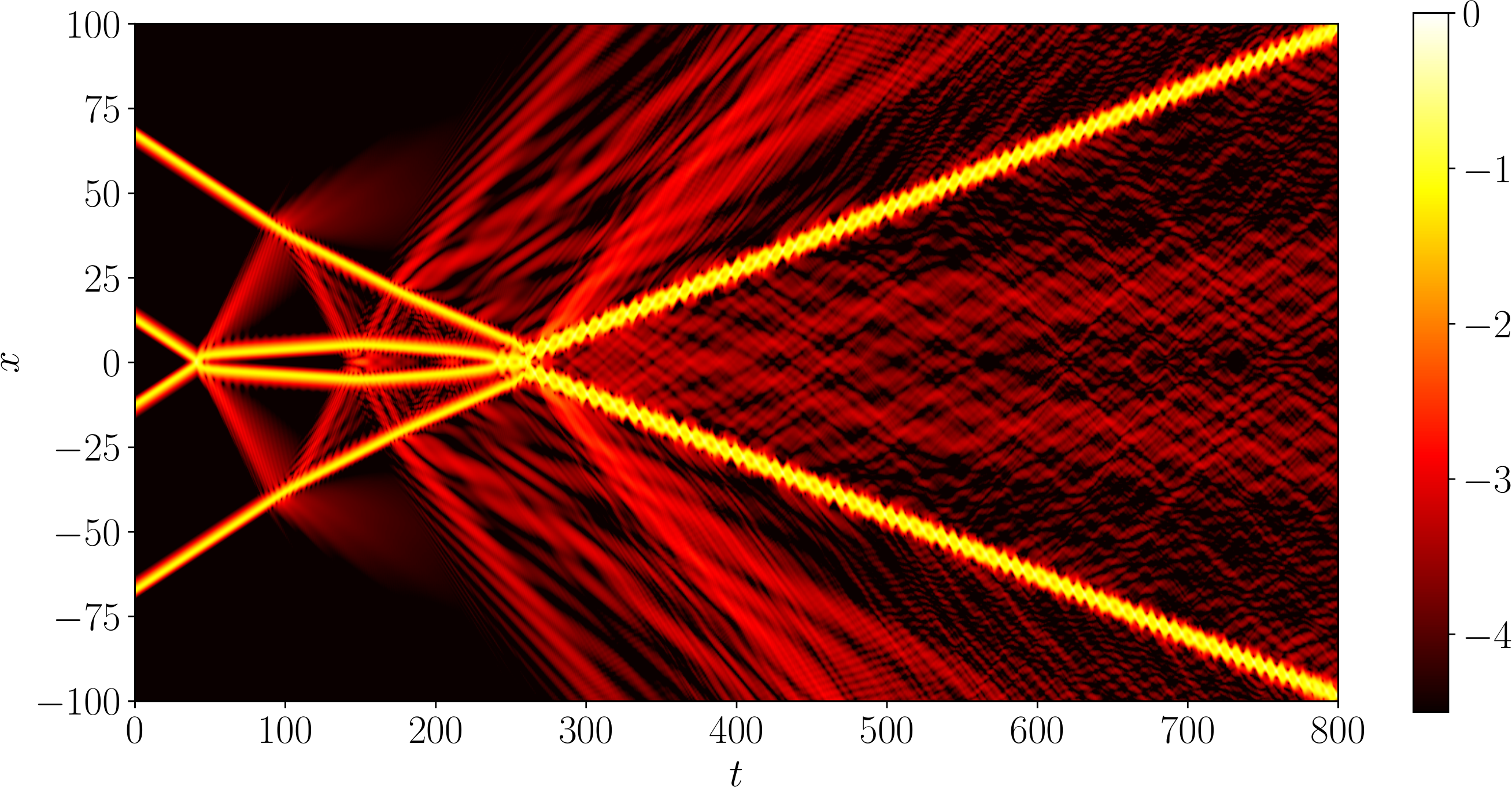}
\caption{The logarithm of energy density for the collision shown in figure \ref{Radiation_effects1}, showing the importance of radiation in this regime.}
\label{Radiation_effects}
\end{figure}

These radiation effects persist away from the $\beta=\infty$ ($\phi^6$) limit and provided $\beta$ is large enough they can have a decisive role in determining the outcome of collision processes. This is illustrated in figures \ref{Radiation_effects1} and \ref{Radiation_effects}. The initial velocity is above the critical velocity of a type I collision, so the first subkink collision produces a pair of subkinks moving away from each other, together with some radiation. This radiation reaches the outer subkinks (which act like tampers) at $t\approx 100$ and reflects back, reversing the direction of motion of the inner kinks and causing them to recollide before they get the chance to interact with the outer subkinks. This time their velocity is less than the type I critical velocity, and therefore they form a bion. Soon after that the outer kinks, slightly slowed down by radiation, collide with this bion and two bions are ejected from the centre, transporting most of the energy away from the central region.

\begin{figure}
\centering
\includegraphics[width=0.49\textwidth]{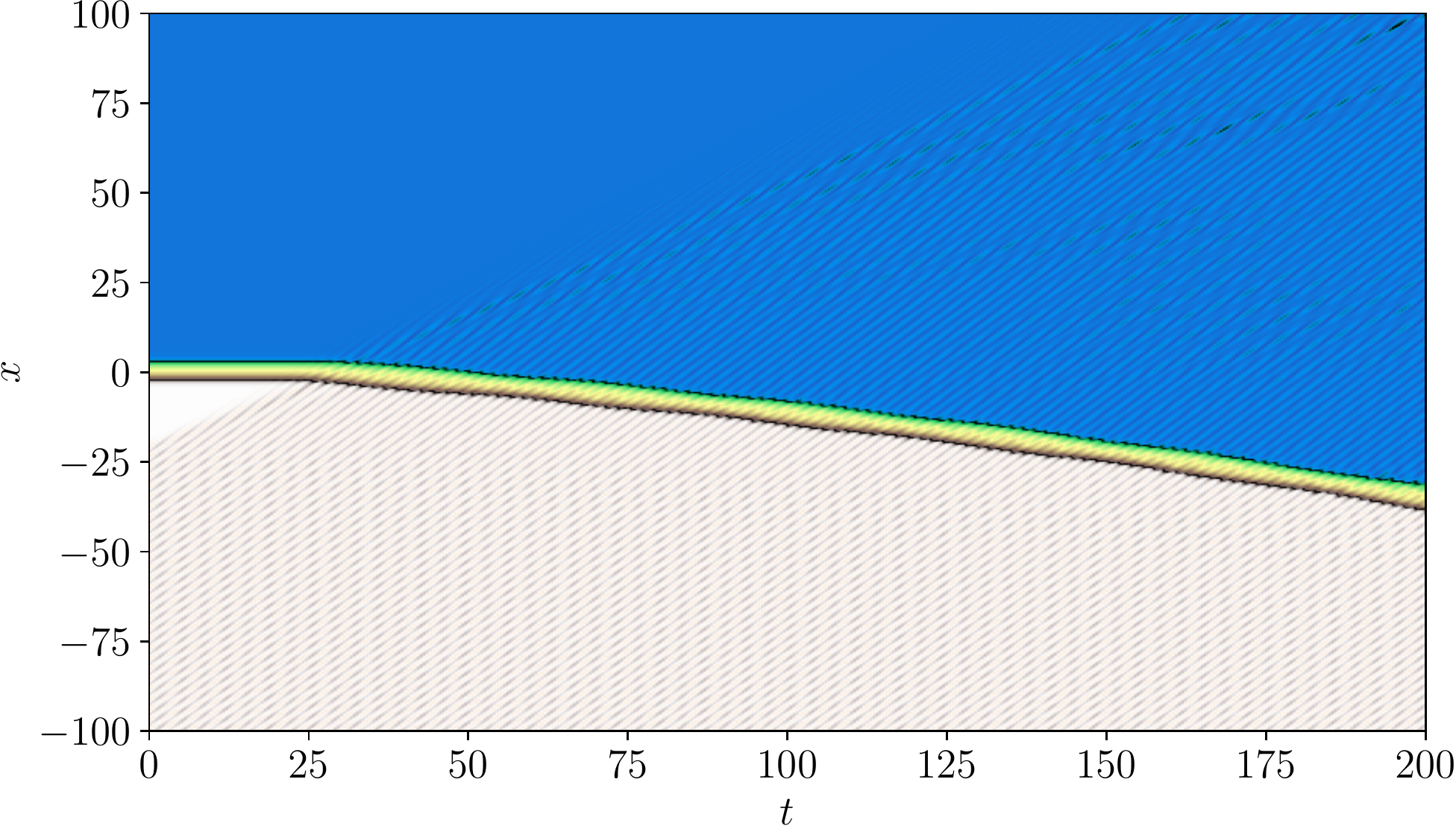}
\includegraphics[width=0.49\textwidth]{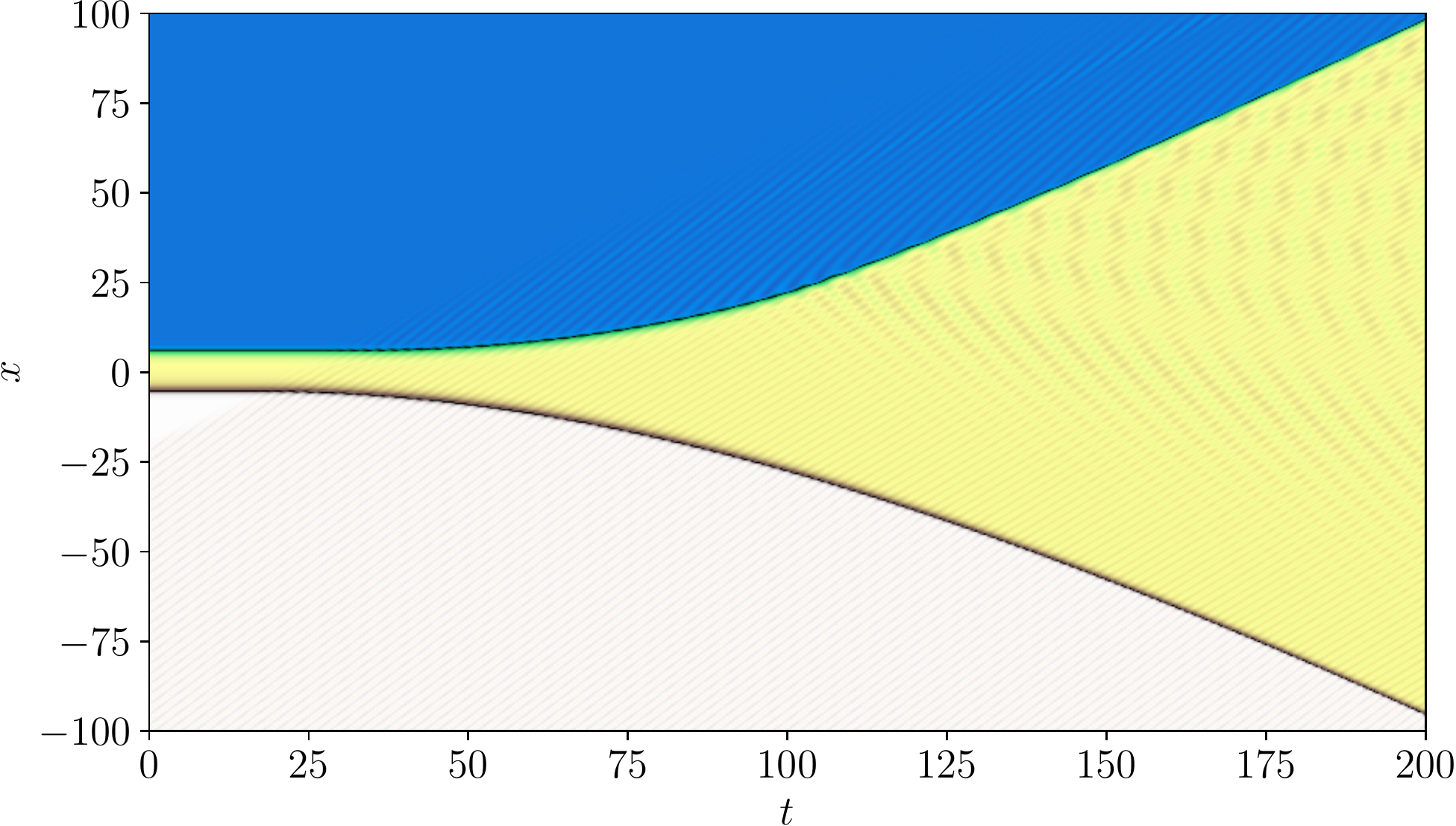}
\caption{
Radiation incident on a single kink for $\beta=3$ (left) and $\beta=5$ (right). In both cases the radiation frequency $\omega=3$; on the left the amplitude $A=0.1$, while on the right $A=0.05$.}
\label{Radiation_effects2}
\end{figure}

Radiation can also have an effect on a single  kink, as illustrated in figure \ref{Radiation_effects2}.
Effectively it acts as a force pushing the subkinks away from each other no matter where the radiation comes from, meaning that an incident burst of radiation can break the subkinks apart, provided $\beta$ is large enough.
Figure \ref{Radiation_effects2} shows the effect of a right-moving wave train with frequency $\omega=3$ and amplitude $A$, 
supported at $t=0$ on $[-\infty,-20)$, on a single kink with $\beta=3$ (left plot) and $\beta=5$ (right plot).
For $\beta=3$ only small oscillations in the kink width could be observed even for the relatively-large amplitude $A=0.1$, though the kink as a whole undergoes a negative radiation pressure. By contrast, for $\beta=5$, which is within the regime where subkinks behave as independent $\phi^6$ kinks, an incident wave train 
with the smaller amplitude $A=0.05$ has no problem  separating the subkinks. 
Over an exponentially-large timescale at least as great as that given by equation (\ref{recollide}) above (which ignores residual radiation pressure), 
the extremely-weak false vacuum pressure will cause these subkinks to recollide, but over the timescales shown
in the plot, the lower (negative spatial coordinates) subkink trajectory is precisely matched by that of the corresponding single $\phi^6$ kink subject to the same incident radiation, as would be expected from our general considerations.

\section{Kink-bion-antikink collisions for $\epsilon=0$}\label{sec4}
As seen in the last section, in the large $\beta$ regime a centrally located
bion formed from a pair of subkinks can play an important role as an intermediate state in collision processes. This bion may then be impacted by other subkinks, either during the initial set of collisions as seen in figure \ref{CollisionTypes}\,b, or in the far future of processes such as those shown in figures \ref{CollisionTypes}\,a and \ref{LastCollision}, where false vacuum pressure will ultimately force the return of the outer subkinks to the central region. These impacts are well modelled by kink-bion-antikink collisions in the $\phi^6$ model. This section is devoted to a study of such processes.  

\begin{figure}
\hspace*{0.02\textwidth}{\small a)}\hspace*{0.48\textwidth}{\small b)}\hspace*{0.41\textwidth}\\
\centering
\includegraphics[width=0.49\textwidth]{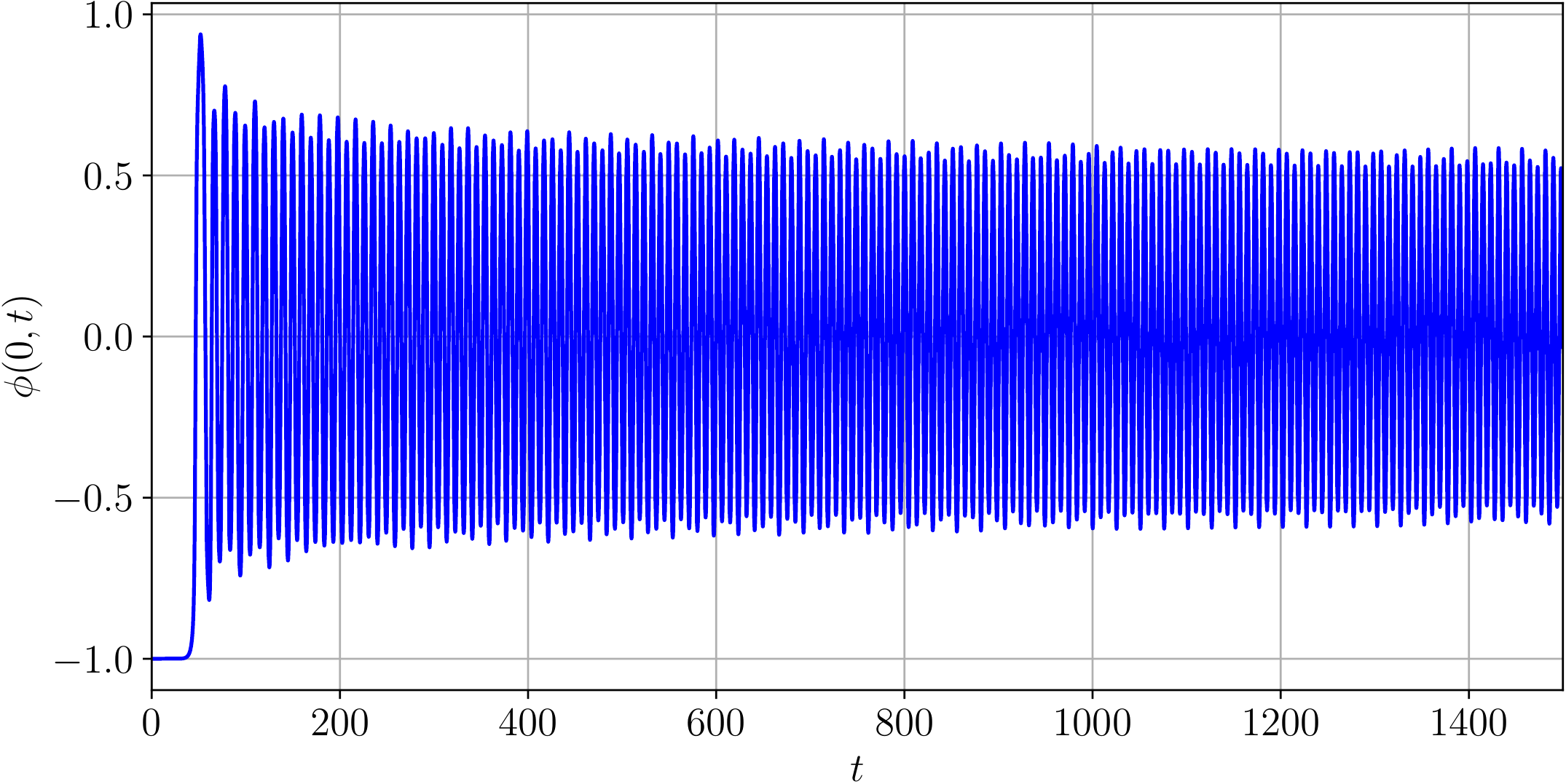}
\includegraphics[width=0.49\textwidth]{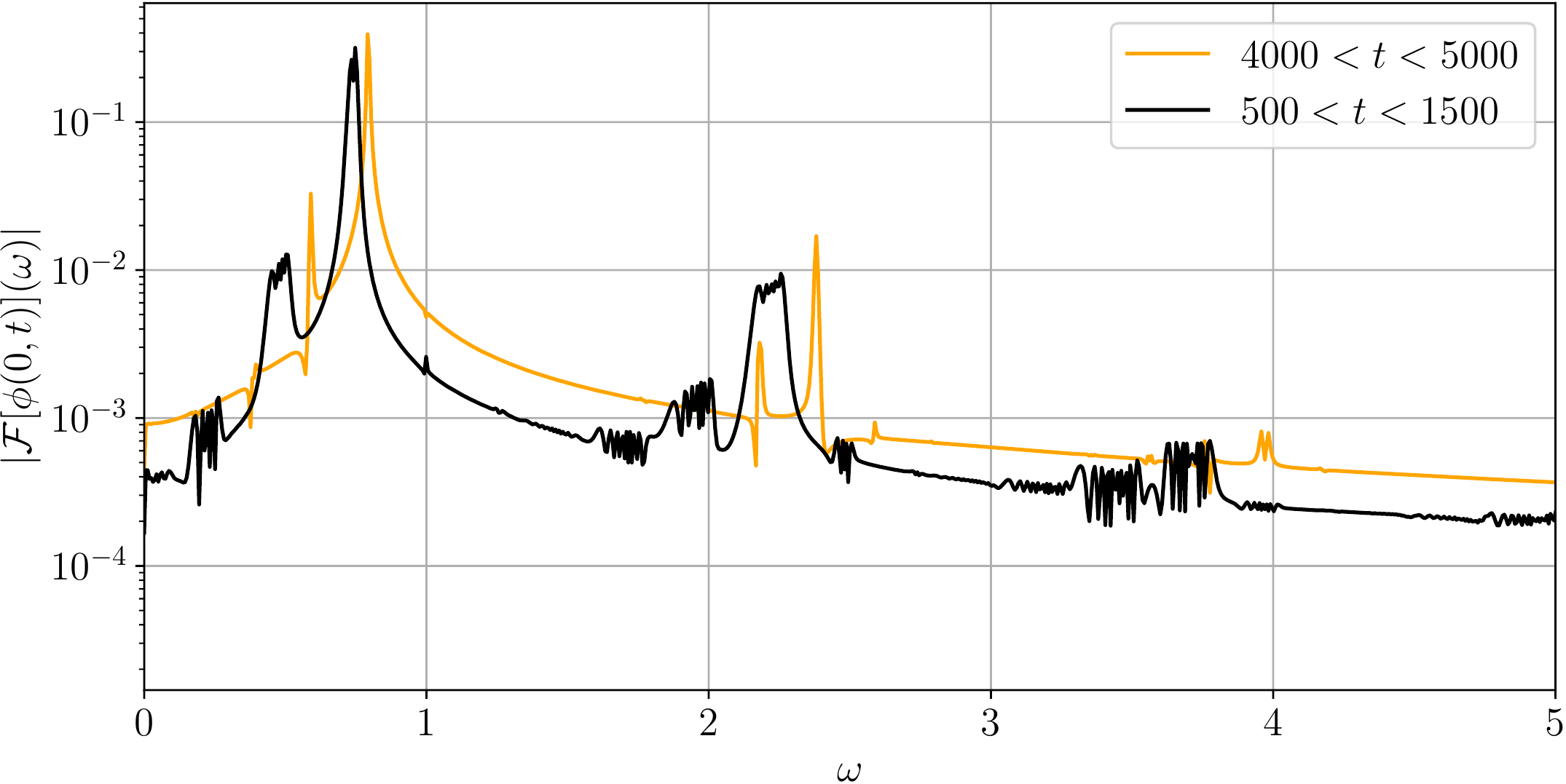}
\caption{Typical bion after the type I $\phi^6$ collision ($v=0.2)$: (a) field at centre, (b) its power spectrum at different times. }
\label{typicalbion}
\end{figure}

Bions formed in the $\phi^6$ model after a type I collision such as that shown in figure \ref{LastCollision2}\,a have surprisingly uniform properties over a wide range of initial velocities, from near zero up to the critical velocity. Their amplitudes do not decay steadily (so they are not pure oscillons\footnote{Some authors reserve the word \textit{oscillon} only to almost periodic solutions in time, in contrast to the quasi-periodic nature of bions created from kink-antikink collisions. In this paper, however, we will use the terms interchangeably.}) but rather have beats which range between $A\sim(0.55-0.65)\pm 0.02$ over the time-scale of the simulation. 
A typical example is shown in figure \ref{typicalbion}. In the spectrum there are three dominating frequencies. At early stages, $500<t<1500$ these are $\omega=0.75$, $0.50$, $2.26=3\times0.75$. At longer times the frequencies change but the rate of change slows down and the peaks are better defined, at $\omega=0.79$, $0.59$, $2.38=3\times0.79$. Note that $2\omega_1-\omega_2\approx 1$ in both cases, where $\omega_1$ is the dominant frequency, and $\omega_2$ corresponds to the highest peak below $\omega_1$. The time between the first and the second bounce in the collision during which the bion is formed increases as the velocity approaches the critical value. But after the second bounce, a bion with properties described above is formed. 

The bions are formed around the $\phi=0$ vacuum and because of the symmetry of the potentials of the Christ-Lee (and $\phi^6$) models they can only have  odd harmonics. Therefore the radiation coming from the first propagating harmonic comes from the third order of nonlinearity and as such is  smaller than the more-usually observed second harmonic radiation (as in the $\phi^4$ model)  making  central $\phi^6$ bions more stable (in the sense that the radiation leading to oscillon decay is relatively weaker).
Note that the potentials around  $\phi=\pm 1$ is not symmetric and therefore oscillons/bions created around those vacua do possess all harmonics, including the even ones.

\begin{figure}
\hspace*{0.02\textwidth}{\small a)}\hspace*{0.48\textwidth}{\small b)}\hspace*{0.41\textwidth}\\
\centering 
\includegraphics[width=0.49\textwidth]{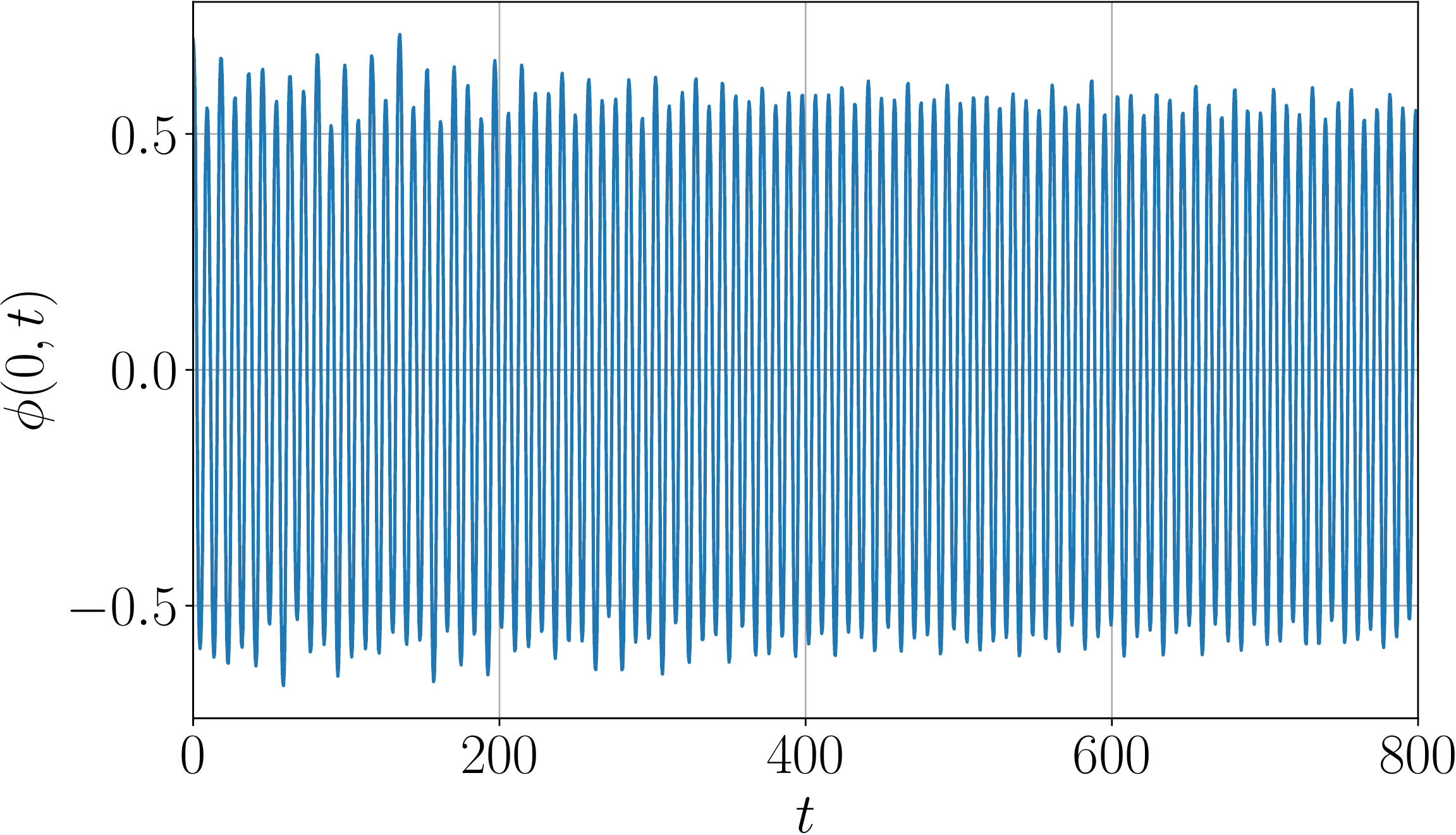}
\includegraphics[width=0.49\textwidth]{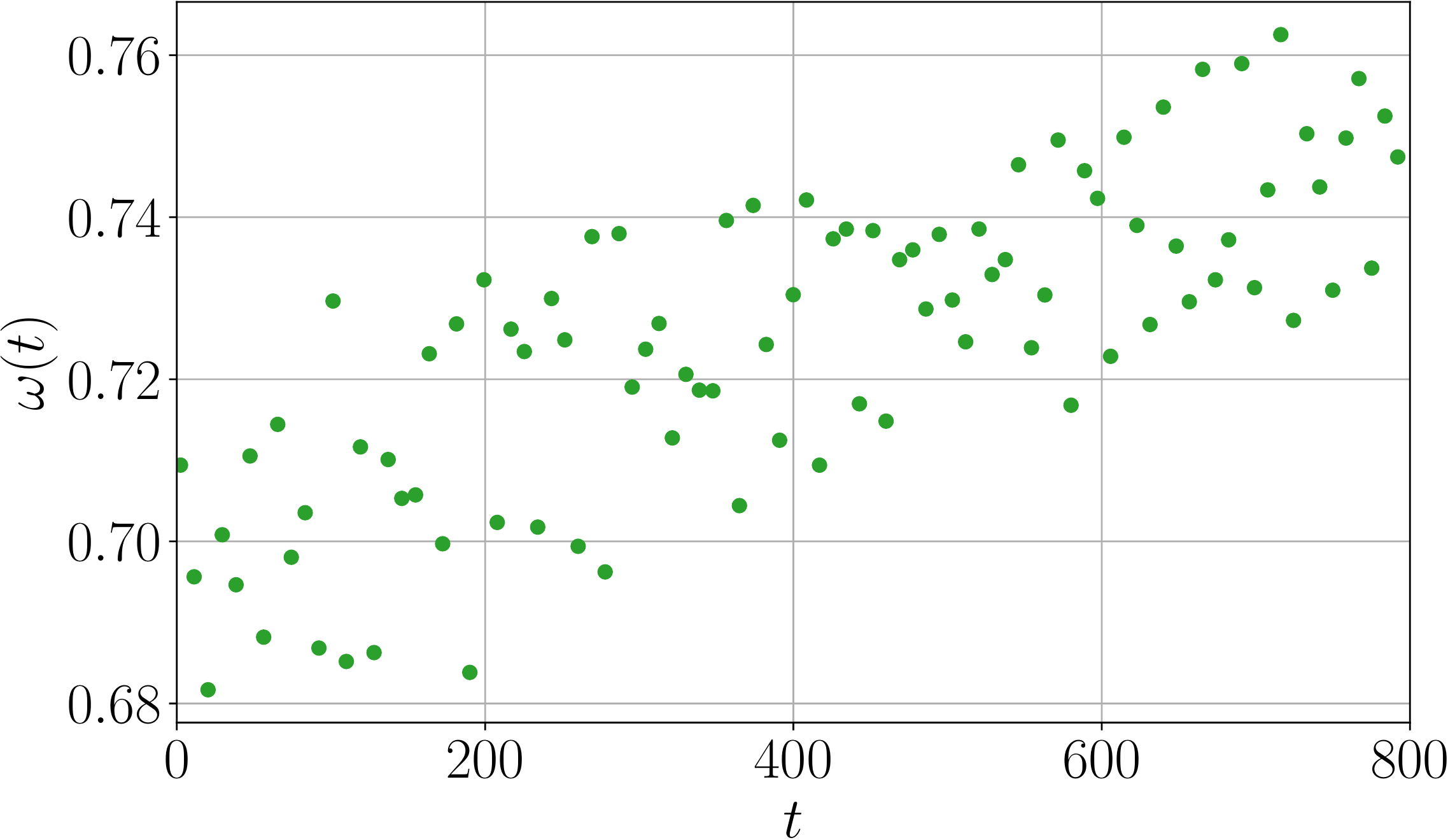}
\caption{Evolution of an oscillon created by gaussian initial conditions. (a) the field at centre, (b) measured frequency defined by time differences between subsequent descending  zeros of $\phi(0, t)$}\label{gaussian_oscillon}
\end{figure}

\begin{figure}

\hspace*{0.08\textwidth}{\small a)}\hspace*{0.48\textwidth}{\small b)}\hspace*{0.41\textwidth}\\
\includegraphics[width=0.49\textwidth]{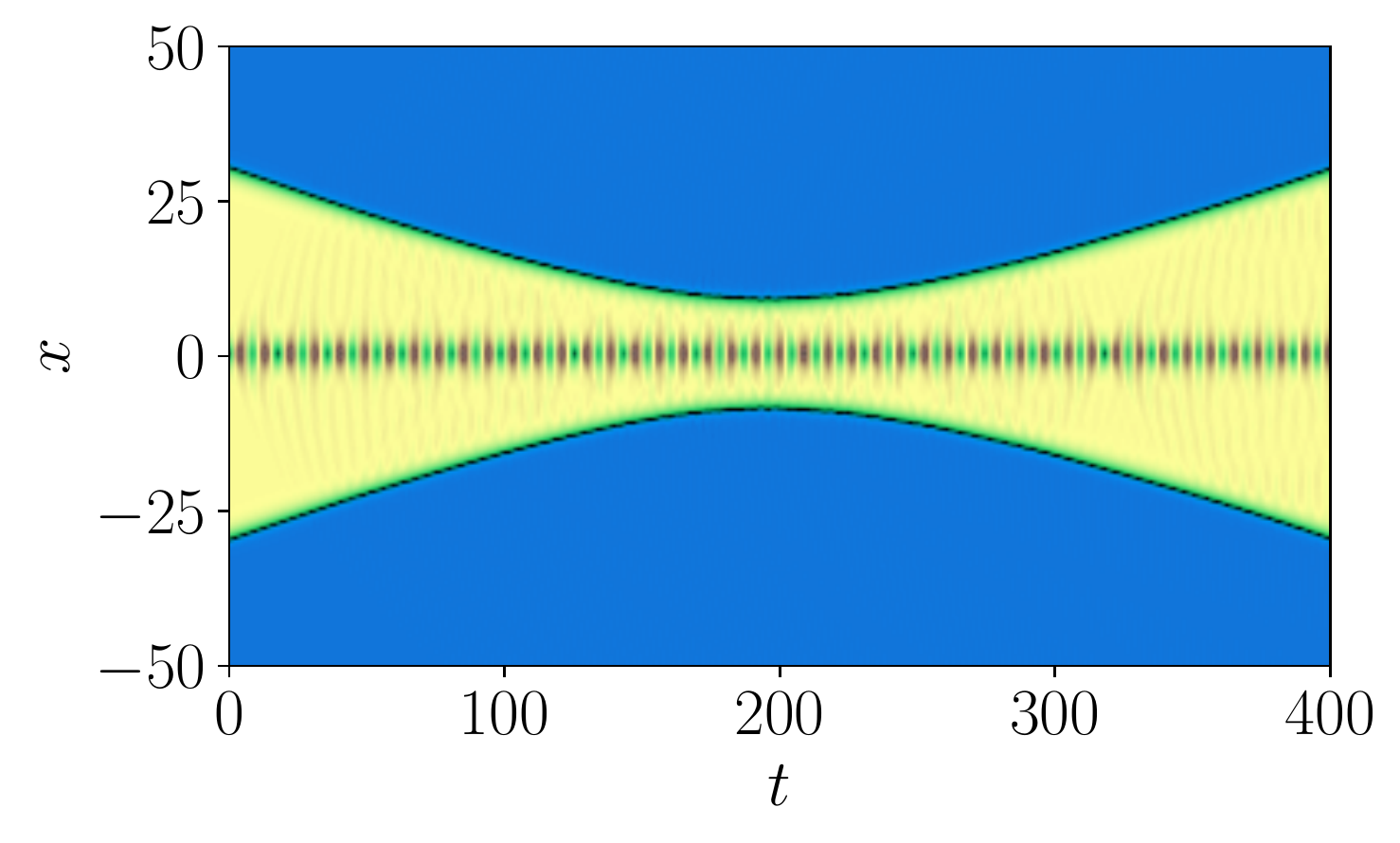}
\includegraphics[width=0.49\textwidth]{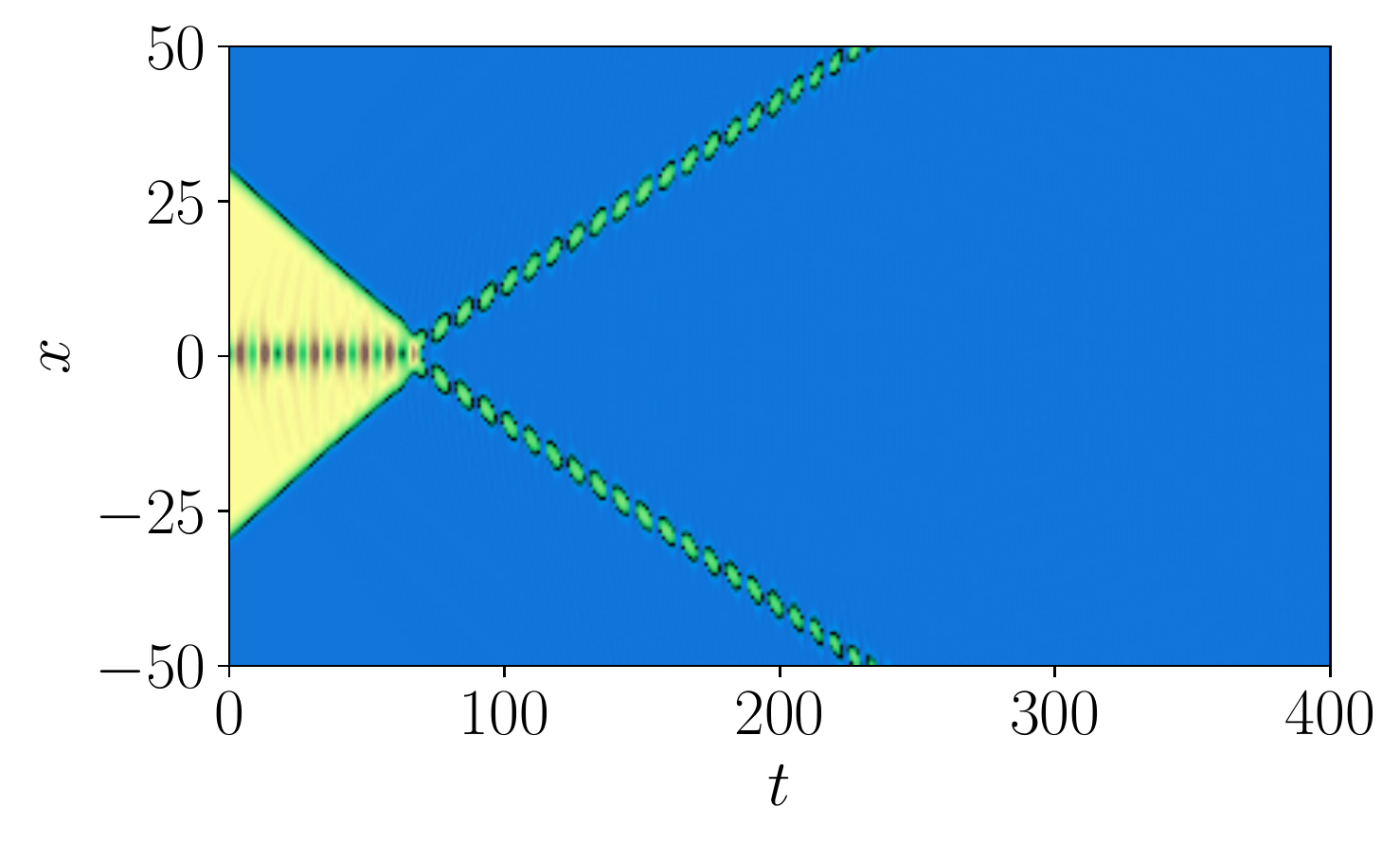}
\hspace*{0.08\textwidth}{\small c)}\hspace*{0.48\textwidth}{\small d)}\hspace*{0.41\textwidth}\\
\includegraphics[width=0.49\textwidth]{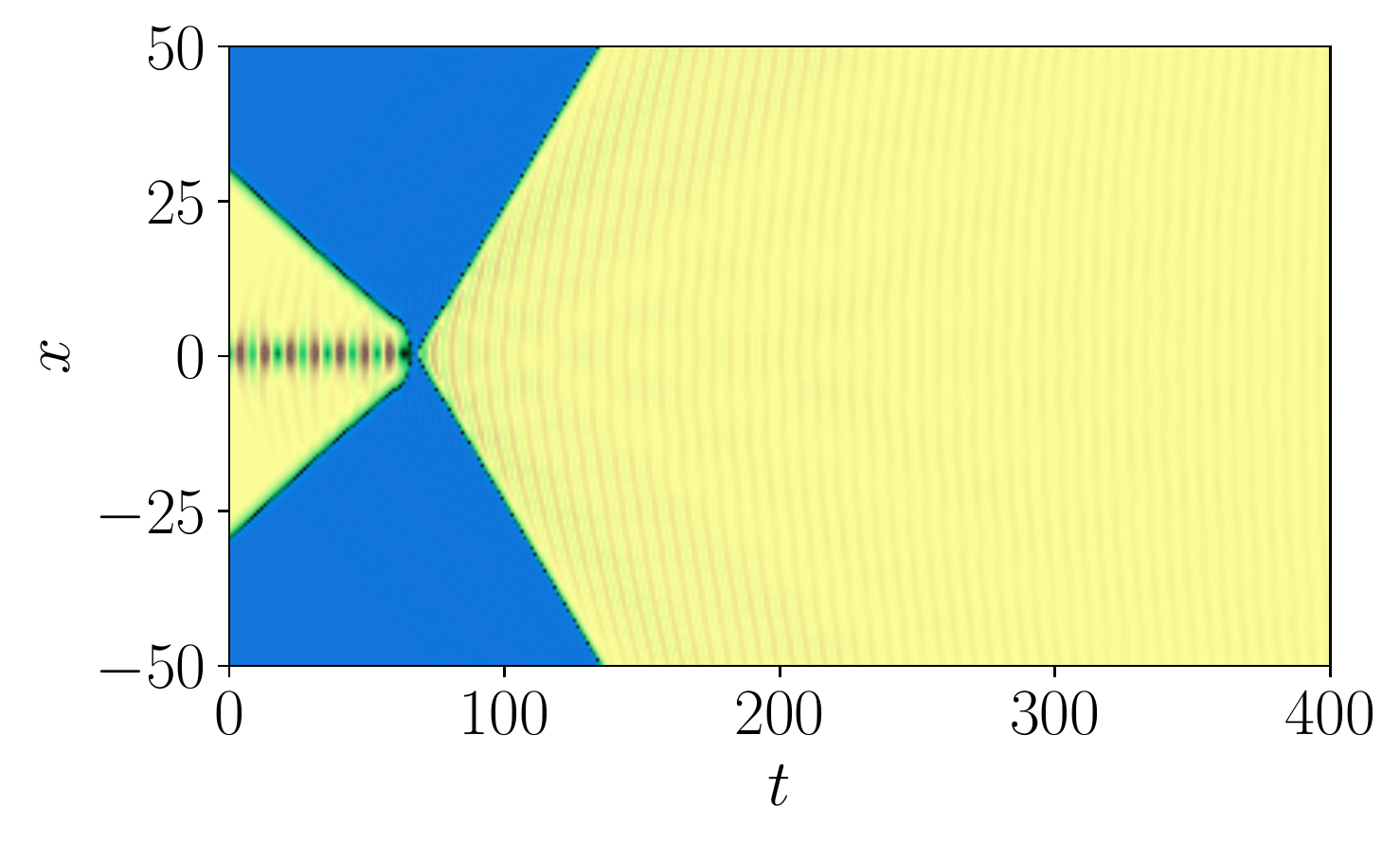}
\includegraphics[width=0.49\textwidth]{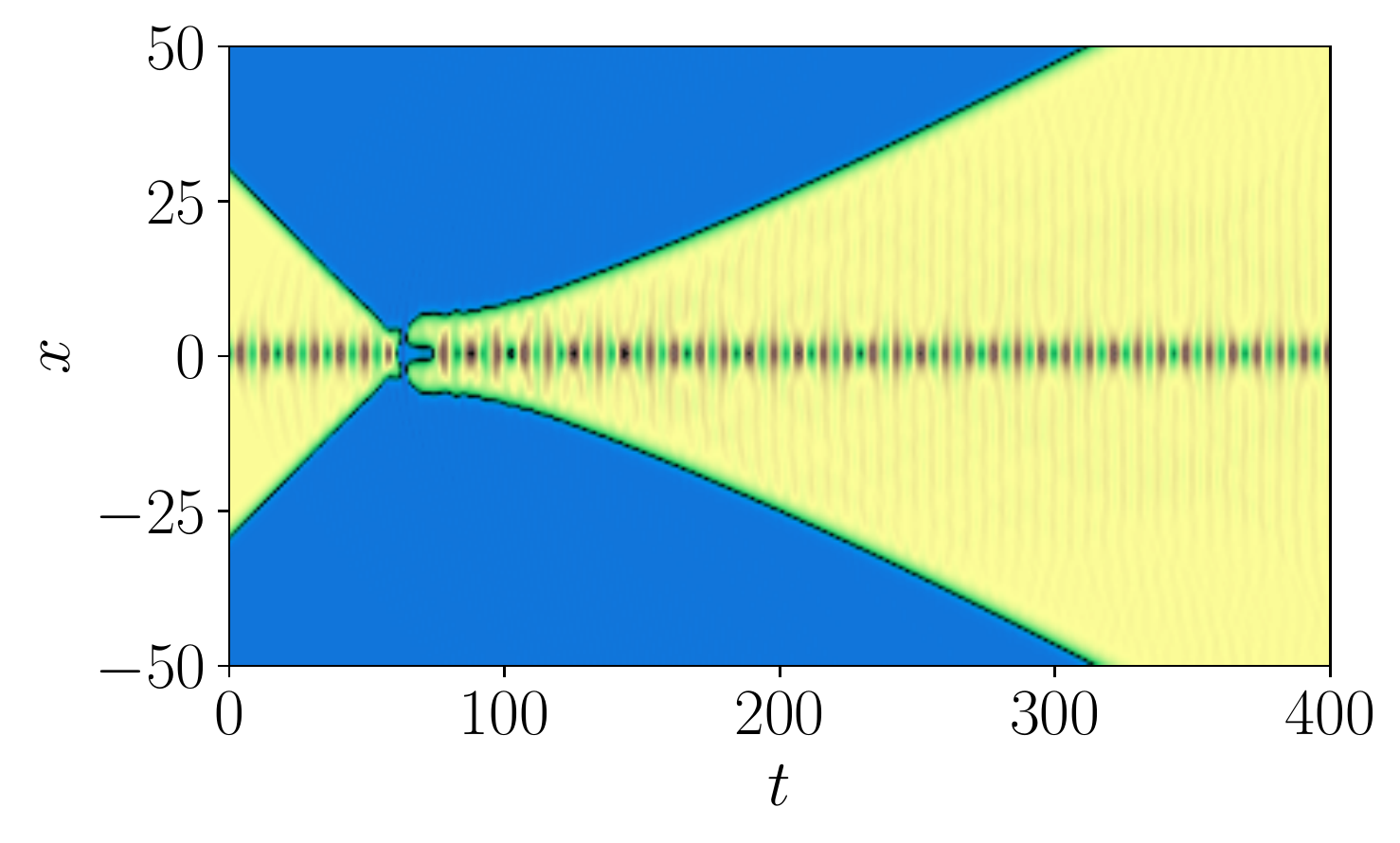}
\centering
\caption{Basic kink-bion-antikink collision scenarios for $x_0=30$. (a) $v_i=0.15$, (b) $v_i=0.40$, (c) $v_i=0.41$, (d) $v_i=0.454$.}
\label{KOAK_scens}
\end{figure}

Collisions between a bion and a pair of symmetrically-placed kinks in the $\phi^6$ model depend strongly on velocity and initial separation of the kinks, but the most important factor is actually the phase of the central bion at the moment of collision. To model this we created a central oscillon  from gaussian initial conditions 
\begin{equation}
    \phi_O(x, 0)=Ae^{-bx^2}\qquad \phi_{O,t}(x, 0)=0.
\end{equation}
Taking $A=0.7$ and $b=0.212$ we were able to create an oscillon with similar properties to the bion resulting from a type I $K \bar K$ collision in the $\phi^6$ model (figure \ref{gaussian_oscillon}). Far away from the oscillon, at positions $\pm x_0$\,, we added $(-1, 0)$ and $(0, -1)$ kinks moving symmetrically towards the centre with initial velocities $\mp v_i$. Then we evolved the system according to the $\phi^6$  ($\epsilon=0$) equation of motion. Some examples of possible solutions are shown  in figure \ref{KOAK_scens}.
For small velocities  (figure \ref{KOAK_scens}\,a) the radiation emitted from the oscillon pushed the kinks away and no contact was made. 
For higher velocities the kinks could split the oscillon and form two escaping bions (figure \ref{KOAK_scens}\,b). Topologically if we can treat the central oscillon as a tightly bound kink -- antikink pair, then after the collision with the incoming kinks different pairs are bound. 
Another possibility is that the kinks bounce off the oscillon, and can either take some of its energy and escape with much higher velocities than $v_i$ (figure \ref{KOAK_scens}\,c) or can lose some of their kinetic energy to excite the oscillon and produce radiation (figure \ref{KOAK_scens}\,d). Note that in the last case the kinks are clearly accelerated by radiation pressure after the interaction.

For each bion there exists a minimal velocity for the incoming kinks below which the collision cannot take place. Assuming a constant radiation force we could expect that the lower critical velocity would be proportional to $\sqrt{x_0}$. 
The measured critical velocities for $x_0=15$ 
and $30$ 
were $0.149$ and $0.186$ respectively which gives the ratio $1.25$ instead of $\sqrt{2}\approx1.41$.
This discrepancy is not surprising, given that in our simulations the force was not constant because the initial gaussian did not have radiation tails instantly.
These tails develop with time. Moreover, the wavefront has a peak which pushes the kinks even more for a short time. Finally, as the solitons slow down so does the radiation frequency observed in the moving frame of the solitons, and the radiation pressure depends on frequency.

\begin{figure}
\hspace*{0.02\textwidth}{\small a)}\hspace*{0.48\textwidth}{\small b)}\hspace*{0.41\textwidth}\\
\centering
\includegraphics[width=0.49\textwidth]{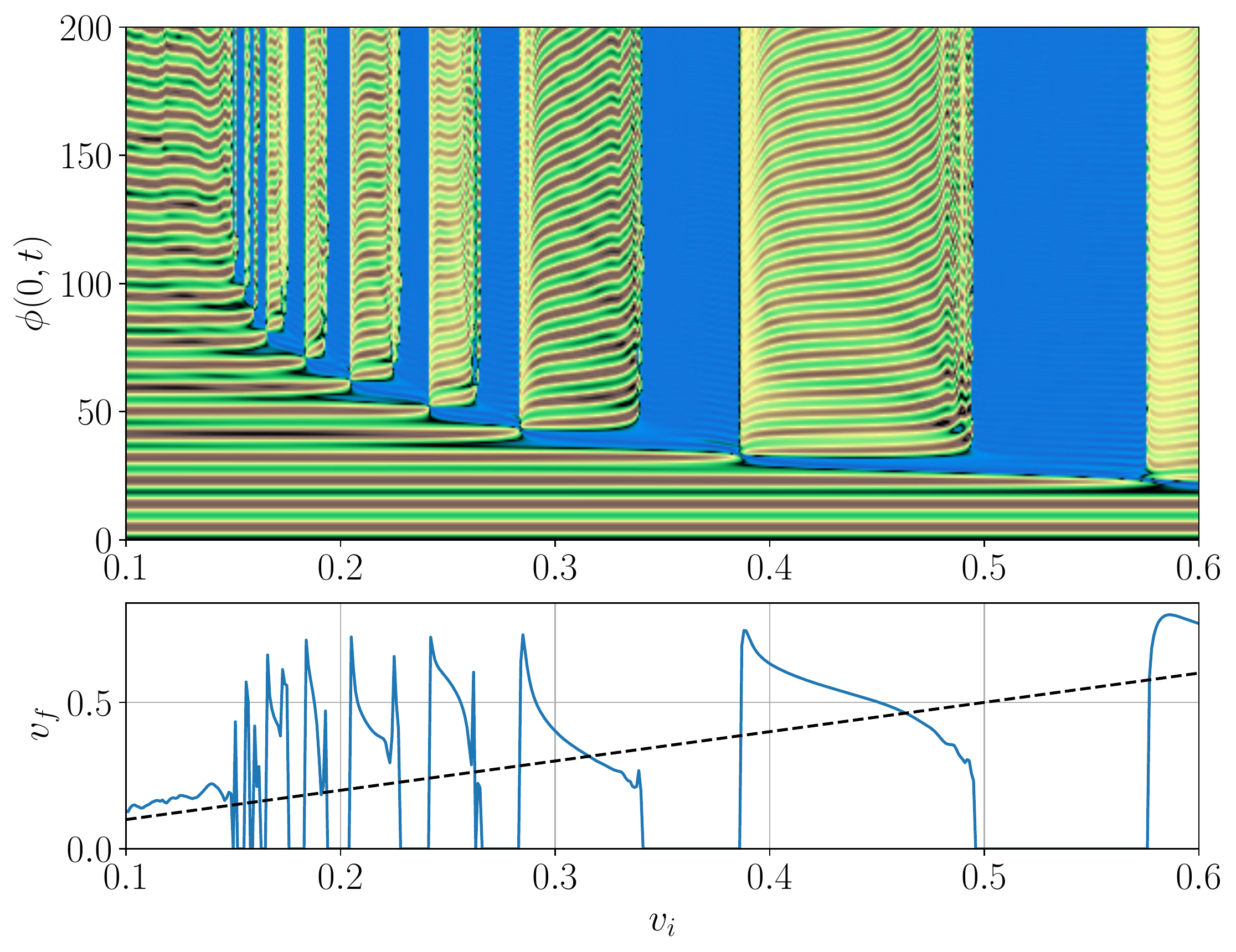}
\includegraphics[width=0.49\textwidth]{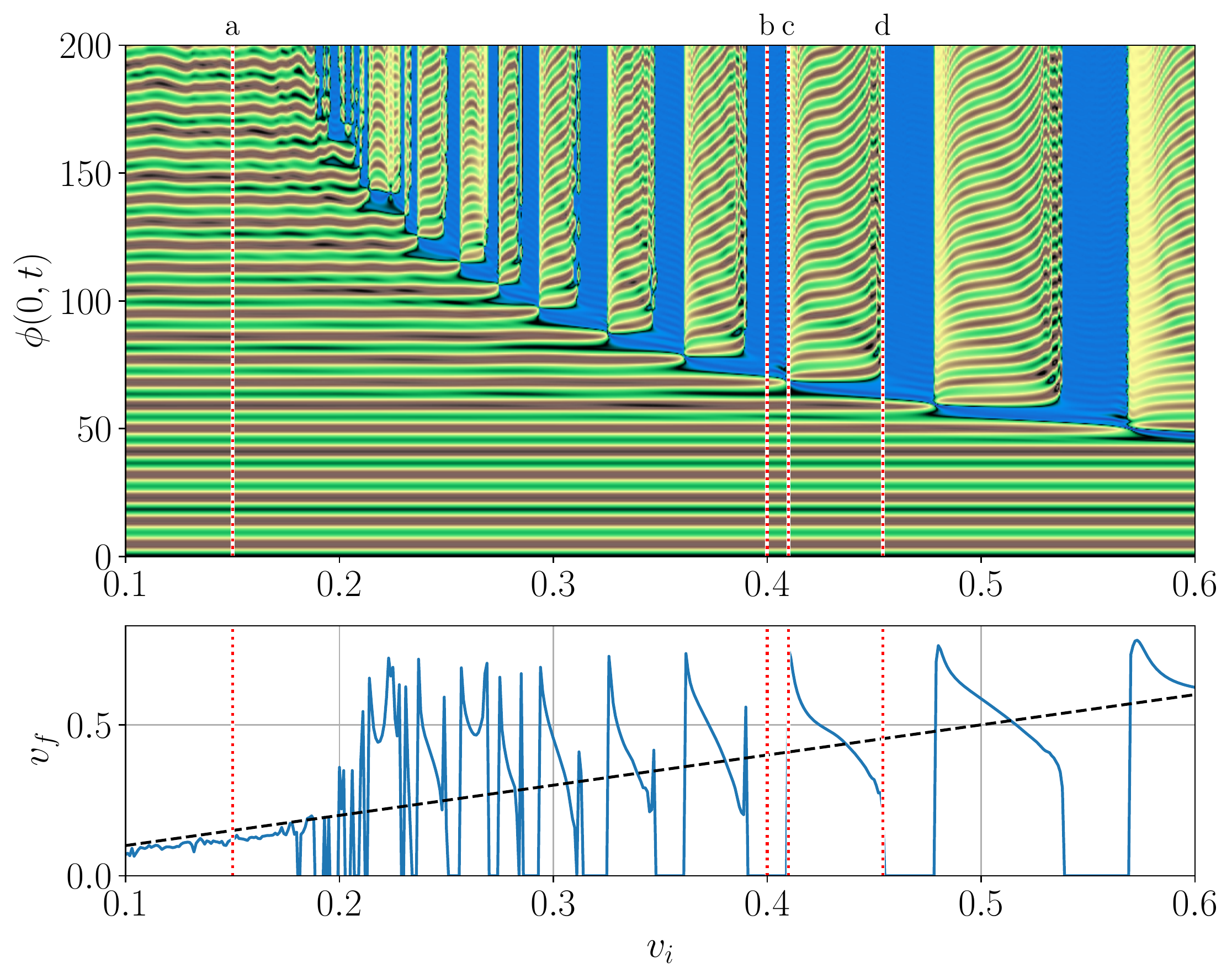}
\caption{Field at centre and escape velocity of a kink for the same central oscillon $A=0.7$ and $b=0.212$ but for different initial separations $x_0=15$ (plot a) and $x_0=30$ (plot b). The vertical dotted lines labelled a-d on plot b indicate the velocities at which the field maps shown in figure \ref{KOAK_scens} were made.}
\label{KOAK_scans}
\end{figure}

In figure \ref{KOAK_scans} we show scans through a range of initial velocities $v_i$. The upper plots show the field value at the centre and the lower plots show the velocities of the escaping kink or subkink. The dashed black line corresponds to $v_i=v_f$, allowing parameters to be found
where the kinks gained speed (scenario \ref{KOAK_scens}\,c).  
Blue columns in the upper plots indicate two escaping bions (scenario \ref{KOAK_scens}\,b).
Green-brown stripes indicate that an oscillon remained in the central region
after the collision which is marked by a blue stripe (scenario \ref{KOAK_scens}\,d).  
Note that there is no blue stripe (collision) below a certain velocity $0.149$ for the left plot with $x_0=15$ and $0.186$ for the right plot with $x_0=30$.
Note that these numbers are consistent with the critical velocities for $x_0 = 15$ and 30 of 0.149 and 0.186 respectively that were quoted earlier.

Just above the velocity when the two escaping bions scenario (figure \ref{KOAK_scens}\,b) changes into kink reflection (figure \ref{KOAK_scens}\,c) there is a significant peak in the final velocity of the kinks. This shows that there is some resonant mechanism of energy exchange between the oscillon and the kinks, which at least initially leads to near-complete destruction of the central oscillon. 

\begin{figure}
\centering
\includegraphics[height=2.22in]{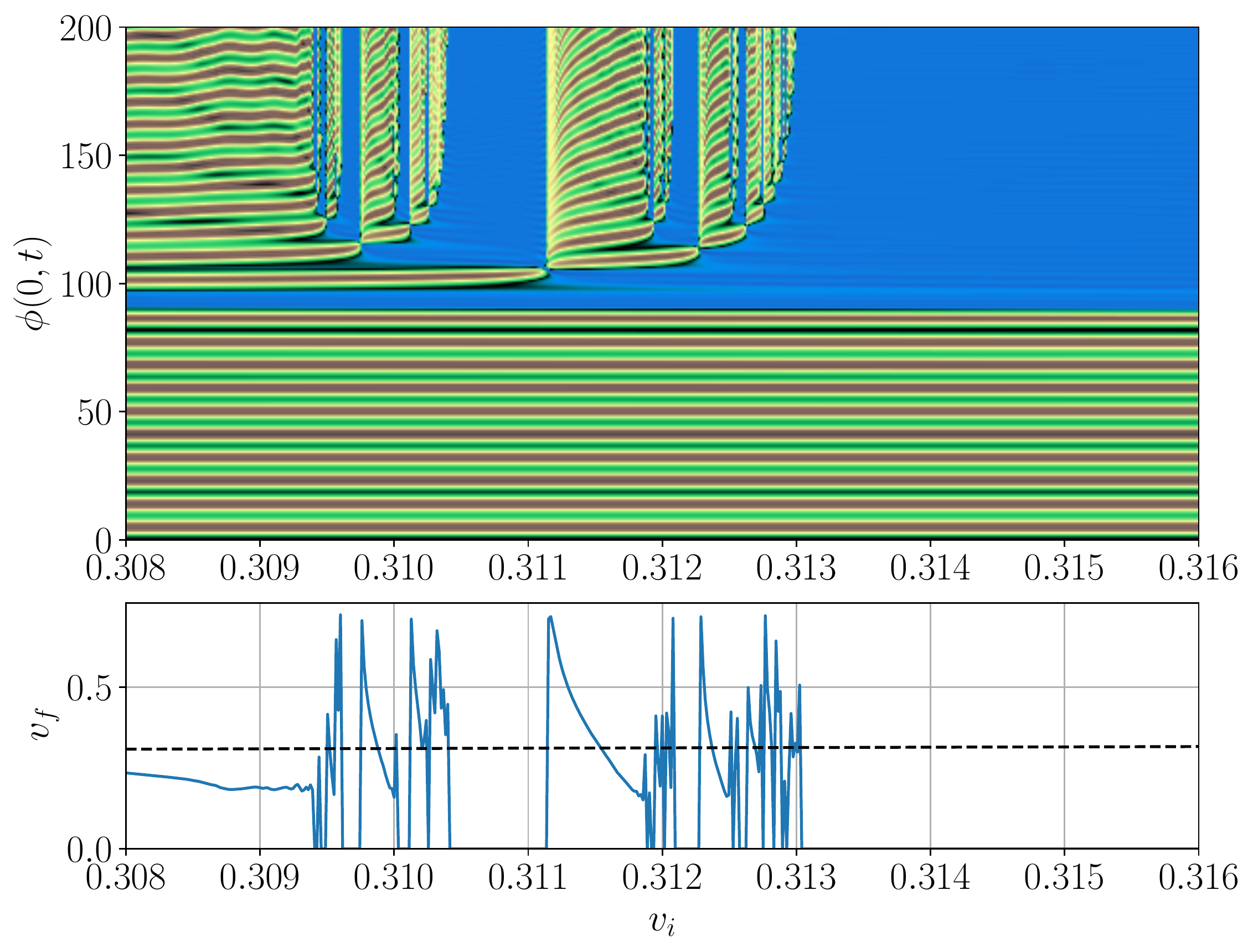}
\includegraphics[height=2.22in]{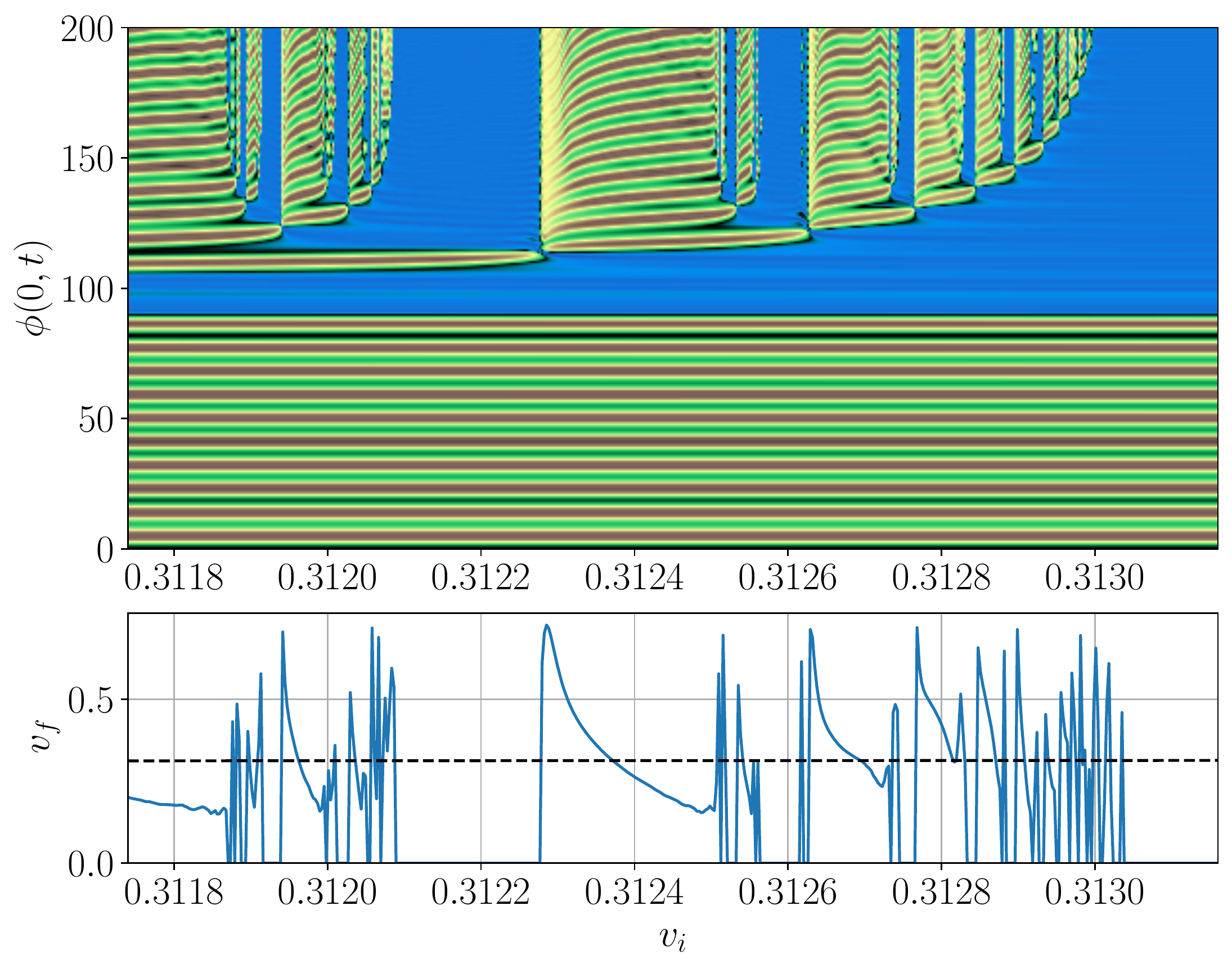}
\caption{Zooms reveal fractal-like features in the kink-bion-antikink collisions for $A=0.7$, $b=0.212$ and $x_0=30$.}
\label{KOAK_zooms}
\end{figure}

\begin{figure}
\hspace*{0.02\textwidth}{\small a)}\hspace*{0.48\textwidth}{\small b)}\hspace*{0.41\textwidth}\\
\centering
\includegraphics[width=0.49\textwidth]{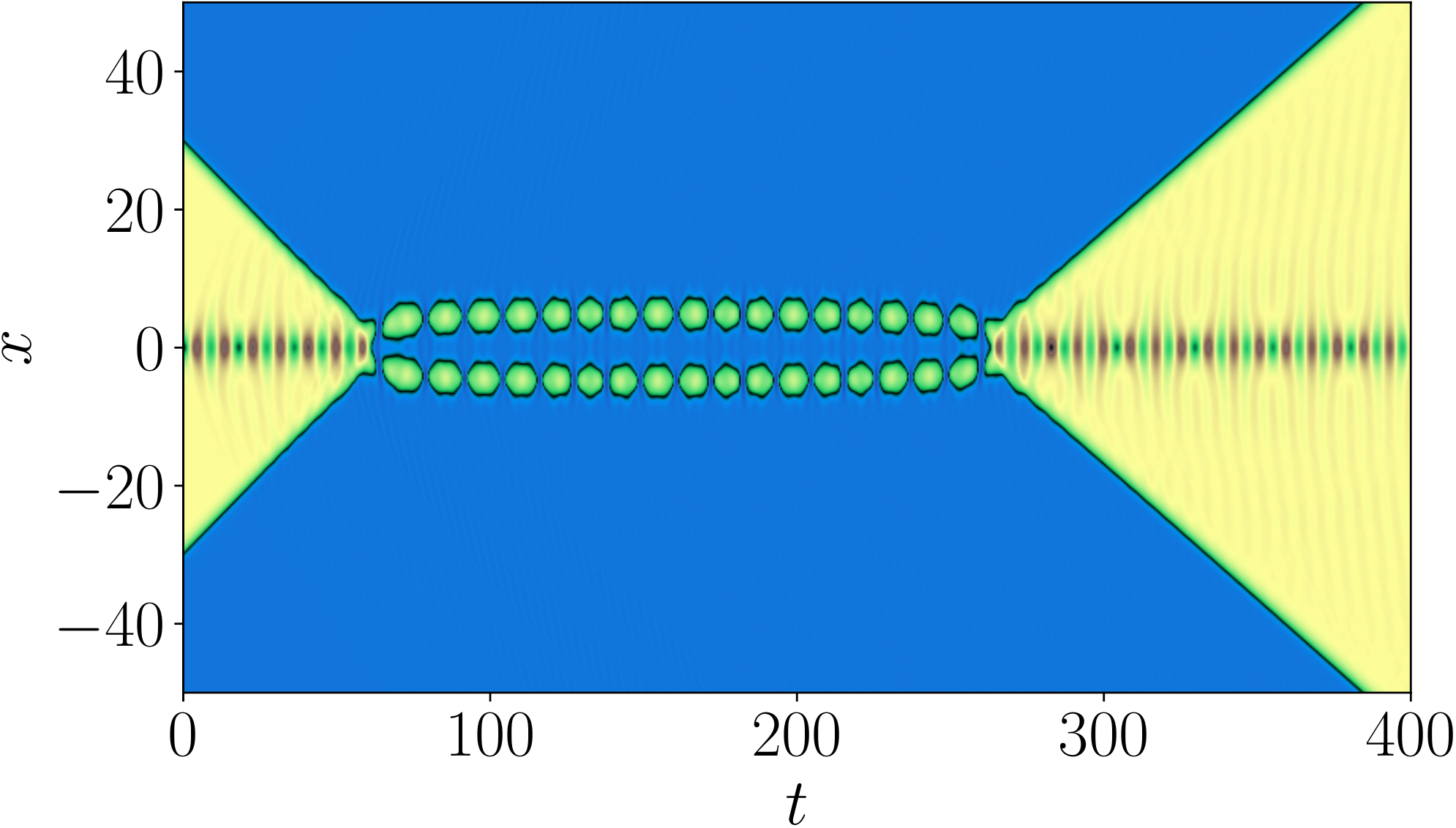}
\includegraphics[width=0.49\textwidth]{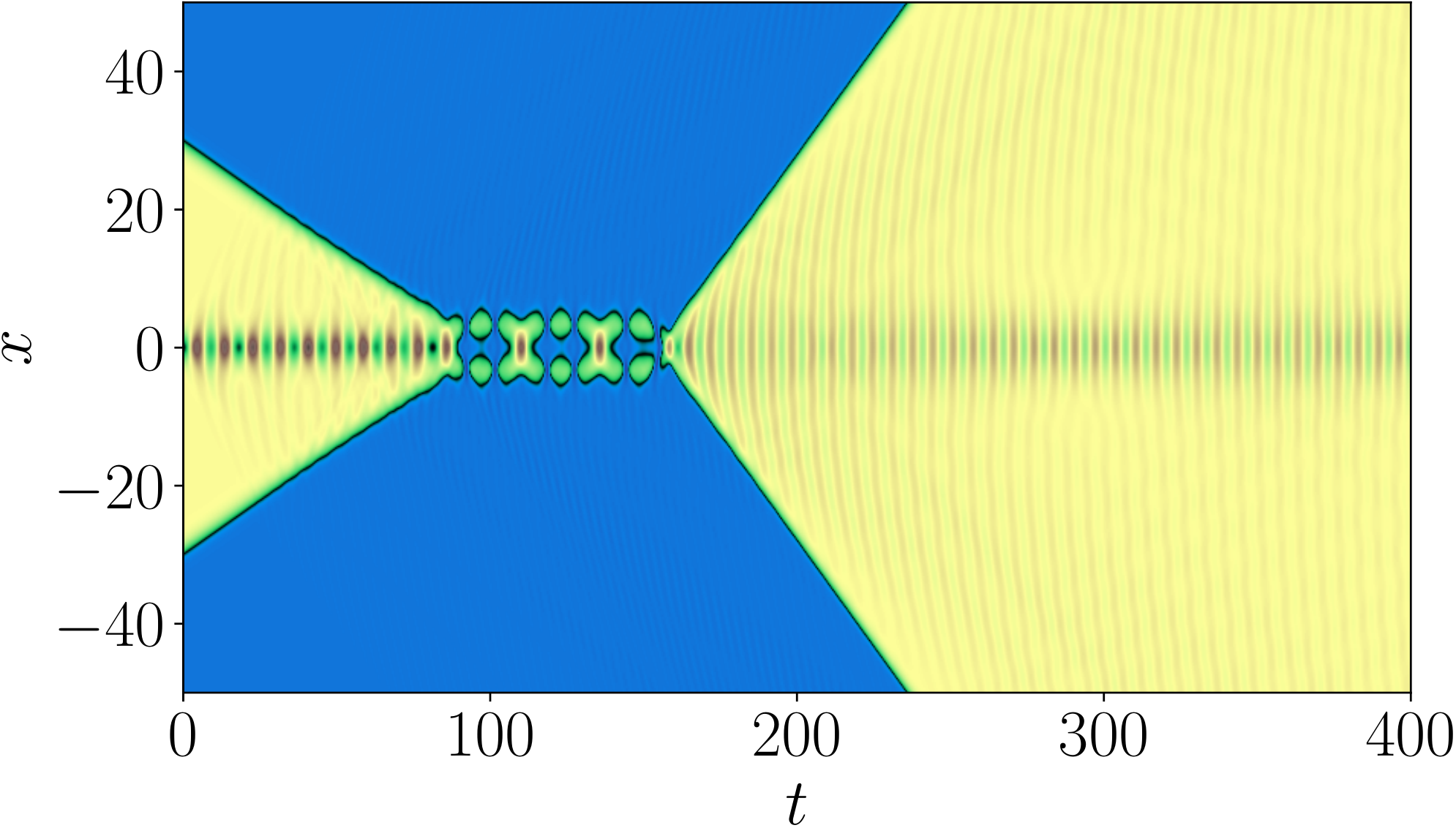}
\caption{Analogues of a long two-bounce window $x_0 = 30$, $v_i=0.4545$ (a),
and a multi-bounce window $x_0= 30$ and $v_i=0.312054$ (b).}
\label{KOAK_zooms2}
\end{figure}

Further zooms near the edges of the green-brown columns in figure \ref{KOAK_scans}, shown in figures \ref{KOAK_zooms}\,a and \ref{KOAK_zooms}\,b, reveal that the structure repeats itself to some extent, exhibiting fractal-like features. In Figure \ref{KOAK_zooms2} we present some typical collision processes where after the initial impact a pair of oscillons is created (corresponding to blue regions at intermediate times in the scans in figures \ref{KOAK_scans} and \ref{KOAK_zooms}) which later annihilates into a kink-antikink pair escaping to infinity and a central oscillon, corresponding to the green-brown columns on the scans. The processes shown in figures \ref{KOAK_zooms2} and \ref{KOAK_scens}\,b, during which incident kinks are converted into oscillons, are reminiscent of the mechanism for resonant scattering of sine-Gordon solitons on a non-integrable boundary discussed in \cite{Arthur:2015mva}, where in some situations an incident kink can be reflected as a breather.

\begin{figure}
\centering
\includegraphics[width=0.75\textwidth]{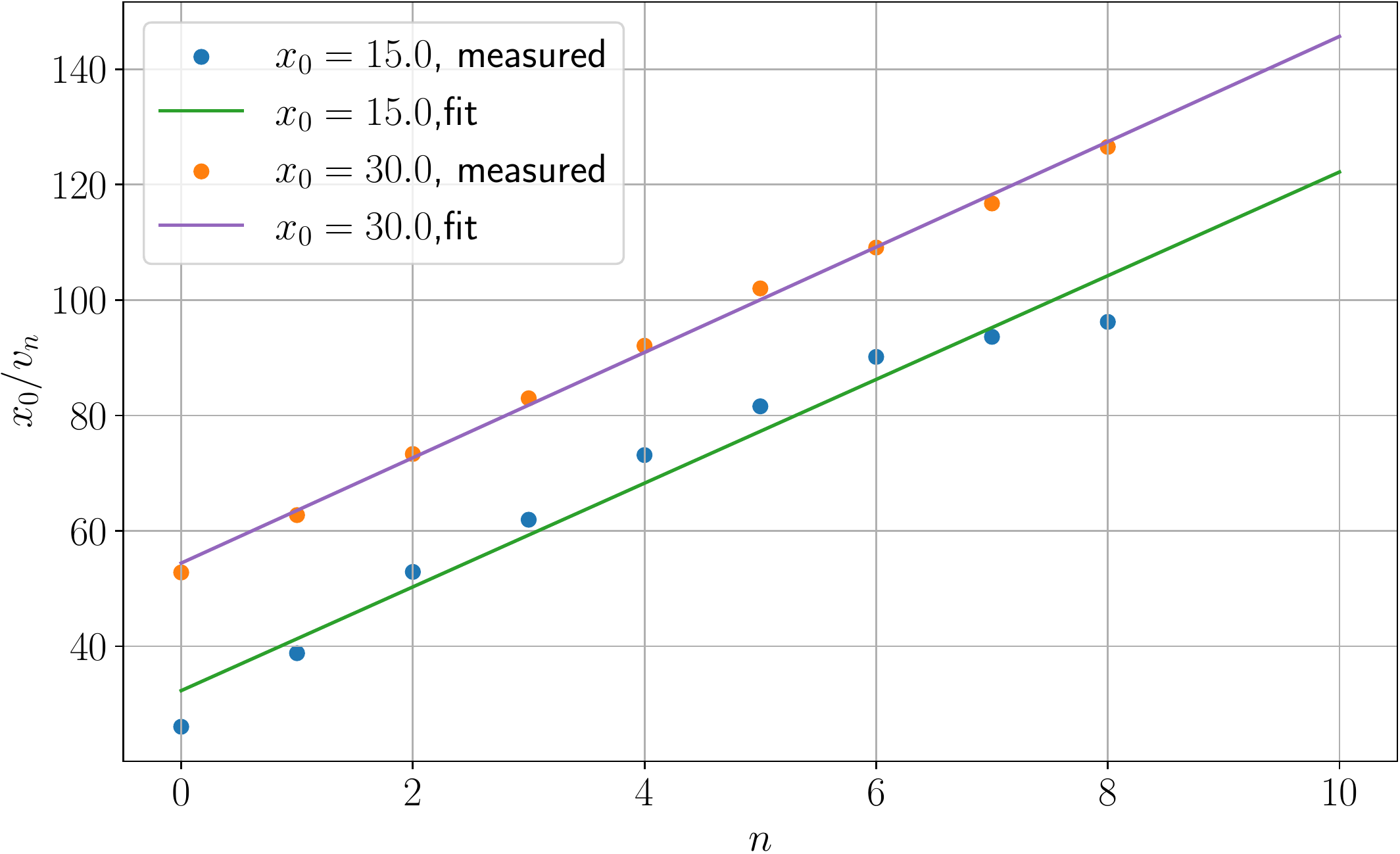}
\caption{
Transition collision time $T_n=x_0/v_n$
as a function of relative window number for $x_0=15$ and $30$, along with the straight line fits.} 
\label{fit_frequencies}
\end{figure}

As already mentioned, a crucial element determining the outcome of a kink-bion-antikink collision is the phase of the central bion.
To verify this we estimate the corresponding resonant frequency from the scattering data, by plotting 
$T_n = x_0/v_n$ \textit{vs} $n$, where $n$ is a relative window number (as counted from higher velocities to lower) and $v_n$ is a measured velocity for 
the transition from the process of ejecting two oscillons (figure \ref{KOAK_scens}\,b) to subkink ejection with maximal velocity (figure \ref{KOAK_scens}\,c).
We expect that 
\begin{equation}
    T_n = \frac{2\pi}{\omega}(n-n_0).
\end{equation}
For initial separations $x_0=15$ and $x_0=30$ the linear fits to the data shown in figure \ref{fit_frequencies} yield
$\omega=0.6992 \pm 0.0340$ and $\omega = 0.6885 \pm 0.0024$. These values correspond very well with the measured oscillon frequencies at small times as shown in figure \ref{gaussian_oscillon}. The  oscillon created has a dominant frequency, but its evolution is not fully periodic, and the measured frequency varies between 0.68 and 0.73.

\section{Conclusions}

The main purpose of this work has been to extend the analysis of the collisional dynamics of  kinks in a 1+1 dimensional scalar theory to the case of weakly bound pairs of kinks. We focused our analysis on the Christ-Lee model, which, depending on the value of the parameter $\beta$, interpolates between the $\phi^4$ model with two vacua ($\beta=0$), and the $\phi^6$ model with a triply-degenerate vacuum ($\beta\to \infty$). 
For sufficiently large values of $\beta$ the  system has a false vacuum at $\phi=0$ and the static solutions of the model have an inner structure as a pair of weakly bound subkinks. In addition to this, the presence of the false vacuum allows for existence of unstable lumps (or sphalerons), which can be produced as intermediate states
in the collisions of the solitons.       

The spectrum of linearized fluctuations about the kinks contains multiple localized modes. Despite this, we observed that the excitation of these modes plays little role in the collisional dynamics of kinks and antikinks
as $\beta\to\infty$. In this limit the subkinks can be treated as independent solitons, and the  collision becomes a multi-step process. Depending on the impact velocity, the first collision of the advanced pair of  subkinks may produce a burst of radiation, a pair of outgoing subkinks or/and an oscillon. In all cases the presence of background radiation affects the subsequent collision of the second pair of  subkinks. This mechanism is related to the appearance of spine structures in the collisions of the solitons. The resulting structure is rather less regular than those observed in previous works, where  internal modes controlled the appearance of the resonance windows. It will therefore
be a challenging question for future work to see whether the spines can be understood within a collective coordinates approach.

\section*{Acknowledgements}

We would like to thank Nick Manton 
for interesting conversations and many insightful remarks.
The research of PED was supported in part by the National Science Foundation under Grant No.\ NSF PHY-1748958, and in part from the STFC under consolidated grant ST/T000708/1; he also wishes to thanks the African Institute for Mathematical Sciences South Africa for hospitality while this work was completed.
TR wishes to thank National Science Centre, grant number 2019/35/B/ST2/00059 and the Priority
Research Area under the program Excellence Initiative – Research
University at the Jagiellonian University in Krak\'ow. YS gratefully acknowledges the support of the Alexander von Humboldt Foundation. 

\bibliographystyle{JHEP}
\bibliography{ref}
\end{document}